\documentclass[prc,nofootinbib,showpacs]{revtex4}

\usepackage{graphicx}
\usepackage[usenames,dvipsnames]{color}
\usepackage{amsmath,amssymb,bbold}
\usepackage{hyperref}
\usepackage{comment}

\begin{document}

\title{Transport coefficients of hot and dense hadron gas in a magnetic field: a relaxation time approach}

\author{Arpan Das$^{1}$}
\email{arpan@prl.res.in}
\author{Hiranmaya Mishra${^1}$}
\email{hm@prl.res.in}
\author{Ranjita K. Mohapatra${^2}$}
\email{ranjita@iitb.ac.in}

\affiliation{${^1}$Theory Division, Physical Research Laboratory,
Navrangpura, Ahmedabad 380009, India,}

\affiliation{${^2}$Department of Physics, Indian Institute of Technology Bombay, Mumbai, 400076, India.}

\begin{abstract}
We estimate various transport coefficients of hot and dense hadronic matter in the presence of magnetic field. 
The estimation is done through solutions of the relativistic Boltzmann transport equation in 
the relaxation time approximation.We have investigated the temperature and 
the baryon chemical potential dependence of these transport coefficients. Explicit calculations are 
done for the hadronic matter in the ambit of hadron resonance gas model.
We estimate thermal conductivity,
electrical conductivity and the shear viscosity of hadronic matter in the presence of a
uniform magnetic field. Magnetic field, in general, makes the transport coefficients anisotropic. It is also observed that all the transport coefficients perpendicular to the magnetic field are smaller compared to their isotropic counterpart.
\end{abstract}

\maketitle

\section{INTRODUCTION}

\label{intro}

  Strongly interacting matter produced in relativistic heavy-ion collision experiments
  at Relativistic Heavy Ion Collider (RHIC) and Large Hadron Collider (LHC) give us a unique opportunity to 
study strong interaction in the nonperturbative regime. For a comprehensive understanding of the hot and 
dense QCD (quantum chromodynamics) medium produced in these experiments, transport coefficients play a crucial
role. Large number of experimental data indicate the formation of quark-gluon plasma in 
high multiplicity heavy-ion collision experiments. Quark gluon plasma produced in the initial stage 
of heavy ion collision shows collective motion, undergoes subsequent space-time evolution and eventually gets 
chemically and the thermally equilibrated and results in a hadronic medium. Hydrodynamical modeling of 
the strongly interacting matter has been routinely used to study the transverse particle spectra of 
hadrons emanating  out of the interaction region. In the context of hydrodynamical modeling, the dissipative effects
 can be important and the related transport coefficients e.g. shear and bulk viscosity etc can play a significant role
in this hydrodynamical evolution. In various literature it has been argued that a small value of shear viscosity
 to entropy density ratio ($\eta/s$) can explain the flow data\cite{HeinzSnellings2013,RomatschkeRomatschke,KSS}. 
One of the remarkable achievements of the viscous hydrodynamical model is the prediction of a small value of $\eta/s$  and perfect fluid behavior of the strongly interacting matter. A small value of $\eta/s$ of the strongly coupled plasma produced in the heavy-ion collision is in accordance with the lower bound (KSS bound) for the same, $\eta/s=\frac{1}{4\pi}$ obtained using gauge gravity duality (AdS/CFT correspondence). Prediction of the small value of  
  shear viscosity to entropy density ratio motivated a large number of investigations in understanding
 the microscopic origin of transport coefficients \cite{KSS}.
  It is important to mention that 
  KSS bound has been derived for a strongly coupled quantum field theory having conformal symmetry. However, QCD is not 
conformal and the deviation of the conformality is encoded in the bulk viscosity $\zeta$, of the medium.
  \cite{gavin1985,kajantie1985, DobadoTorres2012,sasakiRedlich2009,sasakiRedlich2010,KarschKharzeevTuchin2008,
FinazzoRougemont2015,WiranataPrakash2009,JeonYaffe1996}. Bulk viscosity encodes the conformal
 measure $(\epsilon-3P)/T^4$ of the system and lattice QCD simulations show a non monotonic
behavior of both $\eta/s$ and $\zeta/s$  near the critical temperature $T_c$.  \cite{DobadoTorres2012,sasakiRedlich2009,sasakiRedlich2010,KarschKharzeevTuchin2008,
FinazzoRougemont2015,WiranataPrakash2009,JeonYaffe1996}. 
Non monotonic behavior of bulk and shear viscosity is perhaps very natural because of the emergent scale, 
in this case, $\Lambda_{QCD}$, near the phase transition region. 

Apart from the production of strongly interacting matter in a heavy-ion collision, generation of a non vanishing 
electromagnetic field in the non central heavy-ion collisions allows one to study some novel interplay of QED 
and QCD interactions. The strength of the magnetic field at the initial stages in these collisions can be 
large, at least of the order of several $m_{\pi}^2$ at RHIC energies and may even be larger i.e. 
of the order of 15 $m_{\pi}^2$ at LHC energies  \cite{mclerran2008,skokov}.
Presence of large external magnetic field can have nontrivial effects on 
the properties of QGP, as well as on the subsequent hadronic medium. This has motivated a large number of 
investigations on the properties of hot and dense matter under strong fields. 
Nontrivial nature of the QCD vacuum along with a strong magnetic field can give rise to novel CP-violating 
effects such as chiral magnetic effect  \cite{kharzeevbook}.
Fluid dynamical behavior of the strongly interacting matter in the presence of 
magnetic field has been investigated within the framework of Magnetohydrodynamic simulations \cite{MHD1,MHDajit}.
To study the phenomenological manifestation of magnetic field on the strongly interacting matter one requires the initial magnetic field to survive for at least a few Fermi proper time.
The transport coefficient which plays the important role for the survival of magnetic field in a plasma is the electrical conductivity. In the magnetohydrodynamic
limit when the electrical conductivity of the medium is infinite, magnetic field is frozen in the plasma \cite{MHD1,MHDajit,TuchinMHD,MoritzGreif,electricalcond1,electricalcond2,electricalcond3,electricalcond4,
electricalcond5,electricalcond6,electricalcond7,electricalcond8,electricalcond9,electricalcond10,electricalcond11,electricalcond12,
electricalcond13,electricalcond14,electricalcond15}.  This apart thermal conductivity also
plays a significant role in the hydrodynamical evolution  \cite{danicol2014,Kapusta2012}.
Various approaches e.g. perturbative QCD, different effective models etc have been used to estimate various transport coefficients of the QCD matter  \cite{danicol2018,PrakashVenu,WiranataPrakash2012,
KapustaChakraborty2011,Toneev2010,Plumari2012,Gorenstein2008,Greiner2012,TiwariSrivastava2012,GhoshMajumder2013,Weise2015,GhoshSarkar2014,
WiranataKoch,WiranataPrakashChakrabarty2012,Wahba2010,Greiner2009,KadamHM2015,Kadam2015,Ghoshijmp2014,Demir2014,Ghosh2014,smash,bamps,bamps2,
urqmd1,GURUHM2015, ranjitahm,amanhm1,amanhm2,arpanhm}. In the presence of constant magnetic field, the transport coefficients
 no longer remain isotropic.
It can be shown that in the presence of magnetic field in general there can be five coefficients for shear viscosity, two coefficients bulk viscosity, and three coefficients for thermal conductivity  \cite{Rischkemag}. In Ref.\cite{Rischkemag}
for dissipative magnetohydrodynamics of strongly interacting medium a complete set of transport coefficients,
 consistent with the Curie and Onsager principles, has been derived for thermal conduction, 
shear viscosity and bulk viscosity. Further using 
Zubarev’s non-equilibrium statistical operator method,
 Kubo formulae for these transport coefficients has been derived in Ref.\cite{Rischkemag}.

  In the present work, we investigate thermal conductivity, electrical conductivity and shear viscosity 
for the hot and dense hadron gas produced in the subsequent evolution of QGP, in heavy-ion collisions,
 in the presence of a magnetic field. In our previous works, we have investigated electrical conductivity 
and Hall conductivity for hot and dense hadronic matter as well as for quark-gluon plasma \cite{hallhrg,hallqgp}. 
It was seen that for strongly interacting medium, the electrical conductivity decreases in the presence of magnetic 
field while the Hall conductivity displays a non monotonic behavior with  magnetic field. We had also pointed out 
that at zero baryon chemical potential Hall conductivity vanishes due to opposite gyration of particles 
and antiparticles. Only for non vanishing baryon chemical potential, Hall conductivity has a finite value 
\cite{semiconductor, pairplasma1,pairplasma2,pairplasma3}. In these investigations, we had only considered the field 
configurations where the electric and the magnetic field are perpendicular 
  to each other. In general, electric and magnetic field can have more general configurations. In the present
 investigation, we have considered a somewhat general configurations of electric and magnetic field. 
For a general configuration of electric and magnetic field, we have calculated all the components of 
thermal conductivity, electrical conductivities as well as shear viscosity. It is important to 
mention that the formalism that we use to calculate shear viscosity and electrical conductivity 
in the presence of magnetic field has been developed in Ref.s\cite{landaubook,tuchin3,sedrakianPRD}. 
  In the present investigation, we study the effect of magnetic field on various transport
 coefficients of the hot and dense hadronic matter in a magnetic field using the hadron resonance gas model 
(HRGM) within the framework of relaxation time approximation (RTA). It ought to be mentioned that a 
large magnetic field produced in the initial stage of heavy-ion collisions can be sustained in the medium with
 finite electrical conductivity. 
  Some transport coefficients, e.g. bulk viscosity, thermal conductivity etc. in the presence of
 magnetic field for the QGP phase using Landau quantization and considering only the lowest Landau 
level have been investigated in Ref. \cite{hattori1,hattori2,hattori3,vinod1}. On the other hand,
 for the hadronic phase considered here, we limit ourselves to the case of small/moderate magnetic fields.
 Naturally, in such a case, while the equilibrium dynamics is decided by strong interaction, the effect 
of magnetic field is reflected through the cyclotron frequency of the individual hadrons. Such an approximation 
has been utilized earlier to estimate transport coefficients \cite{tuchin3,feng2017}. We will follow a similar approach for hadron resonance gas and naturally, the quantum effects due to Landau quantization is not included here.

 The hadronic phase at chemical freeze-out 
 can be described by well celebrated hadron resonance gas (HRG) model \cite{HRG1,HRG2}. 
 To explain the experimental results of the thermal abundance of different hadron ratios in the heavy-ion collisions, HRG model has been successfully used \cite{HRG3}. Assuming a single chemical freezeout surface of strange and non strange particles, HRG model can be described using only two parameters temperature ($T$) and baryon chemical potential ($\mu_B$). Using S-matrix calculation it has been shown that in the presence of narrow-resonances, the thermodynamics of interacting gas of hadrons can be approximated by ideal gas of hadrons and its resonances \cite{HRG4,HRG5}. Information of interaction among different hadrons has been encoded as the resonances. 
Due to this very simple description, HRG model has been well explored regarding
thermodynamics \cite{thermodynamicsHRG1,thermodynamicsHRG2}, conserved charge fluctuations\cite{hrgfluc3,hrgfluc4,rmfluc,hrgfluc5,hrgfluc6,
hrgfluc7} as well as transport coefficients for
hadronic matter\cite{MoritzGreif,PrakashVenu,WiranataPrakash2012,
KapustaChakraborty2011,Toneev2010,Plumari2012,Gorenstein2008,Greiner2012,TiwariSrivastava2012,GhoshMajumder2013,Weise2015,GhoshSarkar2014,
WiranataKoch,WiranataPrakashChakrabarty2012,Wahba2010,Greiner2009,KadamHM2015,Kadam2015,Ghoshijmp2014,Demir2014,Ghosh2014,smash,
bamps,bamps2,electricalcond2,electricalcond3,Plumari2012,urqmd1,GURUHM2015}. Although many improvements have been done on ideal HRG model of non interacting hadrons and its resonances e.g. including excluded volume HRG model \cite{stockerRischke,GURUHM2015} etc, in this investigation we confine ourselves to ideal
HRG model for the estimation of various transport coefficients.

This paper is organized in the following manner. In Sec.\eqref{thermalcondc}
we discuss the formalism of thermal conductivity in the presence of a magnetic field. In Sec.\eqref{electcondc} and Sec.\eqref{shearvisco} we summarize the formalism to estimate electrical conductivity and shear viscosity in the presence of magnetic field.
In Sec.\eqref{HRGmodel} we discuss salient feature of HRG model and summarize the formalism to calculate thermal averaged relaxation time. Then in Sec.\eqref{results} we present the results for thermal conductivity, electrical conductivity and shear viscosity in the presence of magnetic field. Finally we conclude our work with an outlook to it.

\section{Thermal conductivity in the presence of magnetic field}
\label{thermalcondc}

One of the important transport coefficients relevant for thermodynamic system with non zero baryon density is the coefficient for
thermal conductivity.
Thermal conduction arises when energy flows relative to the baryonic enthalpy. 
Heat current of hadron resonance gas in the presence of conserved baryon current can be defined as \cite{gavin1985} ,

\begin{align}
 \mathcal{I}^i = \sum_{a} T_a^{0i}-\frac{\omega}{n_B}\sum_{a}b_a j_{B_a}^i.
 \label{equ1new}
\end{align}
Here $a$ is the particle index, $b_a$ is the baryon number of different particles, e.g. for mesons $b_{\text{meson}}=0$, for baryons $b_{\text{baryon}}=1$ and for antibaryons $b_{\text{antibaryon}}=-1$. $\omega$ is the enthalpy of the system $\omega = \mathcal{E}+P$, $\mathcal{E}$ is the energy density of the system and $P$ is the pressure of the system. In Eq.\eqref{equ1new} $T^{\mu\nu}$ is the energy momentum tensor, $j_{B}^{\mu}$ is the conserved baryon current and $n_B$ is the net number density of baryons. Using the standard definition of $T^{\mu\nu}$ and $j_B^{\mu}$ heat current $\mathcal{I}^i$ can be expressed as, 
\begin{align}
 \mathcal{I}^i & = \sum_a \int \frac{d^3p_a}{(2\pi)^3}p_a^if_a -\frac{\omega}{n_B} \sum_a b_a \int\frac{d^3p_a}{(2\pi)^3} v_a^i f_a\nonumber\\
 & = \sum_a \int \frac{d^3p_a}{(2\pi)^3}\frac{p_a^i}{\epsilon_a}\left(\epsilon_a-b_a\frac{\omega}{n_B}\right)\delta f_a,
\end{align}
here $\epsilon_a=\sqrt{\vec{p_a}^2+m_a^2}$ is the single particle dispersion relation and $\delta f_a=f_a-f_{a_0}$, denotes deviation from equilibrium distribution function $f_{a_0}$. To estimate thermal current $\mathcal{I}^i$ one needs to calculate the deviation $\delta f_a$. 
We start with the relativistic Boltzmann transport equation (RBTE) to estimate $\delta f^a$. The RBTE in presence of magnetic field of a single hadron species  is given by \cite{feng2017}, 

\begin{align}
 \vec{v}_a.\frac{\partial f_a}{\partial \vec{r}}+q_a\left(\vec{v}_a\times\vec{B}\right).\frac{\partial f_a}{\partial\vec{p}_a} = \mathcal{C}[f_a],
 \label{equ1}
\end{align}
where $q_a$ is the electric charge of the particle ``a'' and $\mathcal{C}[f_a]$ is the collision integral. In global thermal equilibrium both L.H.S and R.H.S in Eq.\eqref{equ1} vanishes. Therefore, we can write the kinetic equation as given in Eq.\eqref{equ1} as an equation for deviation from equilibrium $\delta f_a =f_a-f_{a_0}$ \cite{tuchin3},

\begin{align}
 \vec{v}_a.\frac{\partial f_{0_a}}{\partial \vec{r}}+q_a\left(\vec{v_a}\times\vec{B}\right).\frac{\partial (\delta f_a)}{\partial\vec{p_a}} = \mathcal{C}[\delta f_a],
 \label{equ2}
\end{align}

In general the collision integral $\mathcal{C}[f_a]$ can be complicated, however in the relaxation time
approximation (RTA) the collision integral in the local rest frame takes simple form and it can be written as, 

\begin{align}
 \mathcal{C}[\delta f_a]\equiv -\frac{\delta f_a}{\tau_a} ,
 \label{equ3}
\end{align}
where, $\tau_a$ is the relaxation time which determines the time scale for the system to relax 
towards the equilibrium state characterized by the distribution function $f_{0_a}$. 
The underlying assumption of the relaxation time approximation is that the system is slightly 
away from equilibrium due to external perturbation and then it relaxes towards 
equilibrium with the time scale $\tau_a$. In relaxation time approximation external perturbation is 
not the dominant scale. In the strongly interacting medium strong interaction is responsible for
 thermalization of the medium and the external electromagnetic field is a perturbation with 
respect to the strong dynamics.
 The equilibrium distribution function satisfies, 

\begin{align}
 \frac{\partial f_{0_a}}{\partial \vec{p}_a}=\vec{v}_a\frac{\partial f_{0_a}}{\partial \epsilon_a},
 ~~ \frac{\partial f_{0_a}}{\partial \epsilon_a}=-\beta f_{0_a}(1\mp f_{0_a}), 
 ~~f_{0_a}=\frac{1}{e^{(\epsilon_a-\vec{p}_a.\vec{u}-b_a\mu_B)/T}\pm1}
 \label{equ4}
\end{align}
 where the single particle energy is $\epsilon_a(p_a)=\sqrt{\vec{p}_a^2+m_a^2}$,  $\mu_B$ is the baryon chemical potential and $\beta=1/T$, is the 
 inverse of temperature. $\vec{v}_a=\vec{p}_a/\epsilon_a$ is the velocity of the particle, $\vec{u}$ is the fluid velocity. In the local rest frame $\vec{u}=0$. $\pm$
 is for fermion and boson respectively.
 In the presence of temperature gradient and magnetic field we can express deviation of distribution function from the equilibrium in the following way \cite{sedrakianPRD},
 
 \begin{align}
  \delta f_a = (\vec{p}_a.~\vec{\Xi})\frac{\partial f_{0_a}}{\partial\epsilon_a},
  \label{equ5}
 \end{align}
with $\vec{\Xi}$ being related to temperature gradient, the magnetic field and in general can be written as,
\begin{align}
 \vec{\Xi}= \alpha \vec{\nabla}T+ b\vec{h} +c (\vec{\nabla}T\times \vec{h}),
 \label{equ6}
\end{align}
where, $\vec{h}=\frac{\vec{B}}{|B|}$, is the direction of the magnetic field. Using $\delta f$ as given in Eq.\eqref{equ5}, Eq.\eqref{equ2} can be expressed as,
\begin{align}
-qB \vec{v}_a.(\vec{\Xi}\times\vec{h})\frac{\partial f_{0_a}}{\partial \epsilon_a}+\vec{v}_a.\vec{\nabla}f_{0_a} = -\frac{\epsilon_a}{\tau_a}(\vec{v}_a.\vec{\Xi})\frac{\partial f_{0_a}}{\partial \epsilon_a}.
\label{equ7}
\end{align}
Here,
\begin{align}
 \vec{\nabla}f_{0_a} = T \frac{\partial f_{0_a}}{\partial\epsilon_a}\bigg[\epsilon_a\vec{\nabla}\left(\frac{1}{T}\right)-b_a\vec{\nabla}\left(\frac{\mu_B}{T}\right)\bigg].
\end{align}
Using Gibbs-Duhem relation for the steady state $\vec{\nabla}P=\omega\frac{\vec{\nabla}T}{T}+n_BT\vec{\nabla}\left(\mu_B/T\right)=0$, we get \cite{gavin1985}, 
\begin{align}
 \vec{\nabla}f_{0_a} = - \frac{\partial f_{0_a}}{\partial\epsilon_a}\bigg(\epsilon_a-b_a\frac{\omega}{n_B}\bigg)\frac{\vec{\nabla}T}{T}.
\label{equ11new}
 \end{align}
It ought to be mentioned that in the presence of magnetic field first law of thermodynamics as well as Gibbs-Duhem 
relation can get modified. However this modification involves magnetization of the system. 
In this present investigation we are not considering magnetization of the system.
Using Eq.\eqref{equ11new} and the representation of $\vec{\Xi}$ as given in Eq.\eqref{equ6}, Eq.\eqref{equ7}, can be expressed as,
\begin{align}
-q_aB\alpha \vec{v}_a.(\vec{\nabla}T\times\vec{h})-q_aBc(\vec{v}_a.\vec{h})(\vec{\nabla}T.\vec{h})+q_aBc(\vec{v}_a.\vec{\nabla}T)  -\bigg(\epsilon_a-b_a\frac{\omega}{n_B}\bigg)\vec{v}_a.\vec{\nabla}\ln T = -\frac{\epsilon_a}{\tau_a}& \bigg[\alpha \vec{v}_a.\vec{\nabla}T+b(\vec{v}_a.\vec{h})\nonumber\\
& +c \vec{v}_a.(\vec{\nabla}T\times\vec{h})\bigg]
\label{equ8}
\end{align}
Comparing coefficients of different tensor structures in Eq.\eqref{equ8} we get,
\begin{align}
 c= \frac{q_aB}{\epsilon_a}\tau_a \alpha \equiv \omega_{c_a}\tau_a \alpha,
 \label{equ9}
\end{align}
\begin{align}
b= (\omega_{c_a}\tau_a)^2 \alpha (\vec{\nabla}T.\vec{h}), 
\label{equ10}
\end{align}
and
\begin{align}
 q_aBc - \frac{\epsilon_a-b_a\frac{\omega}{n_B}}{T}=-\frac{\epsilon_a \alpha}{\tau_a}.
 \label{equ11}
\end{align}
Here $\omega_{c_a}=\frac{q_aB}{\epsilon_a}$ is the cyclotron frequency of the particle with electric charge $q_a$. Using Eq.\eqref{equ9}, Eq.\eqref{equ10} and Eq.\eqref{equ11} it can be shown that,
\begin{align}
 \alpha = \bigg(\frac{\tau_a}{\epsilon_a}\bigg)\bigg(\frac{\epsilon_a-b_a\frac{\omega}{n_B}}{T}\bigg)\frac{1}{1+(\omega_{c_a}\tau_a)^2}.
\label{equ12}
 \end{align}

Hence the deviation of distribution function from equilibrium is,
\begin{align}
 \delta f_a = \frac{\tau_a (\epsilon_a-b_a\frac{\omega}{n_B})}{1+(\omega_{c_a}\tau_a)^2}\bigg[\vec{v}_a.\vec{\nabla}\ln T+(\omega_{c_a}\tau_a)\vec{v}_a.(\vec{\nabla}\ln T\times\vec{h})+(\omega_{c_a}\tau_a)^2(\vec{v}_a.\vec{h})(\vec{\nabla}\ln T.\vec{h})\bigg]\frac{\partial f_{0_a}}{\partial\epsilon_a}.
\label{equ13}
 \end{align}
With the deviation $\delta f^a$ for the distribution function as above known from RBTE, the heat current as given in Eq.\eqref{equ2} can be expressed as,
  \begin{align}
  \mathcal{I}^i & = \sum_a \frac{\tau_a}{T}\int d^3p_a \frac{p_a^ip_a^j}{\epsilon_a^2}\frac{\left(\epsilon_a-b_a\frac{\omega}{n_B}\right)^2}{1+(\omega_{c_a}\tau_a)^2}\bigg(\nabla^jT+(\omega_{c_a}\tau_a)\epsilon^{jkl}\nabla^kTh^l+(\omega_{c_a}\tau_a)^2h^jh^l\nabla^lT\bigg)\frac{\partial f_{0_a}}{\partial\epsilon_a}\nonumber\\
  & = \sum_a \frac{\tau_a}{3T}\int d^3p_a \frac{p_a^2}{\epsilon_a^2}\frac{\left(\epsilon_a-b_a\frac{\omega}{n_B}\right)^2}{1+(\omega_{c_a}\tau_a)^2}\bigg(\delta^{ik}-(\omega_{c_a}\tau_a)\epsilon^{ilk}h^l+(\omega_{c_a}\tau_a)^2h^ih^k\bigg)\nabla^lT\frac{\partial f_{0_a}}{\partial\epsilon_a}\nonumber\\
  & = - \left(k_0\delta^{ik}-k_1\epsilon^{ilk}h^l+k_2h^ih^k\right)\nabla^kT,
  \label{equ18new}
 \end{align}
here we have introduced the different components of the thermal conductivity $k_0,k_1$ and $k_2$. In Boltzmann approximation these coefficients are explicitly given as respectively,   
   
   \begin{align}
    k_0 = \sum_{a}\frac{g_a}{3T^2}\int \frac{d^3p_a}{(2\pi)^3} \bigg(\frac{p_a^2}{\epsilon_a^2}\bigg)\frac{(\epsilon_a-b_a\frac{\omega}{n_B})^2}{1+(\omega_{c_a}\tau_a)^2}f_{0_a}\tau_a,
    \label{equ19}
   \end{align}
   \begin{align}
    k_1 = \sum_{a}\frac{g_a}{3T^2}\int \frac{d^3p_a}{(2\pi)^3} \bigg(\frac{p_a^2}{\epsilon_a^2}\bigg)\frac{(\epsilon_a-b_a\frac{\omega}{n_B})^2(\omega_{c_a}\tau_a)}{1+(\omega_{c_a}\tau_a)^2}f_{0_a}\tau_a,
    \label{equ20}
   \end{align}
      \begin{align}
    k_2 = \sum_{a} \frac{g_a}{3T^2}\int \frac{d^3p_a}{(2\pi)^3} \bigg(\frac{p_a^2}{\epsilon_a^2}\bigg)\frac{(\epsilon_a-b_a\frac{w}{n_B})^2(\omega_{c_a}\tau_a)^2}{1+(\omega_{c_a}\tau_a)^2}f_{0_a}\tau_a.
    \label{equ21}
   \end{align}
Here $\epsilon_a$, $g_a$, $\tau_a$ are the single particle dispersion relation, degeneracy factor and relaxation time of ``a-th'' particle species. From Eq.\eqref{equ19} it is clear that for non vanishing magnetic field thermal conductivity gets modified. It may be noted that in the absence of the magnetic field, the coefficients $k_1$ and $k_2$ vanishes and the thermal conductivity becomes isotropic and is given by the coefficient $k_0$. Further $k_2$ and $k_1$ are associated with $\vec{\nabla}T.\vec{h}$ and $\vec{\nabla}T\times \vec{h}$ terms in the expression of $\delta f_a$. Hence if $\vec{\nabla}T$ is perpendicular to $\vec{B}$ then $k_2$ vanishes and there are two non vanishing coefficients of the thermal conductivity $k_0$ and $k_1$. For a general configuration
of temperature gradient and magnetic field 
 $\vec{\nabla}T. \vec{h}$ can be non zero, hence in that case all the three components $k_0$, $k_1$ and $k_2$ are non vanishing.
 We may point out that in the absence of magnetic field the expression of $k_0$ as given in Eq.\eqref{equ19} can also be derived systematically as shown in Ref.\cite{albrightkapusta,paramitahm2016} in the Landau frame where the flow velocity is identified with the energy flow rather than baryon number flow as in the Eckert frame.

\section{Electrical conductivity in the presence of magnetic field}
\label{electcondc}
Similar to thermal conductivity, to calculate electrical conductivity we start with the relativistic Boltzmann transport equation (RBTE) of single hadron species in the presence of external electromagnetic field  \cite{feng2017}, 

\begin{align}
 p^{\mu}\partial_{\mu}f+qF^{\mu\nu}p_{\nu}\frac{\partial f}{\partial p^{\mu}}=\mathcal{C}[f],
 \label{equ22new}
\end{align}
here $q$ is the electric charge of the particle, $p^{\mu}$ is the particle four momenta, $F^{\mu\nu}$ is the electromagnetic field strength tensor and $\mathcal{C}[f]$ is the collision integral. For a static and homogeneous case, in relaxation time approximation we can write the kinetic equation as given in Eq.\eqref{equ22new} as an equation for deviation from equilibrium $\delta f =f-f_0$ \cite{tuchin3},

\begin{align}
q \vec{E}.\frac{\partial f_0}{\partial \vec{p}}+q\left(\vec{v}\times\vec{B}\right).\frac{\partial (\delta f)}{\partial\vec{p}} = \mathcal{C}[\delta f]\equiv -\frac{\delta f}{\tau},
 \label{equ23}
\end{align}
In the presence of electric and magnetic field we can
take an ansatz for the deviation of the equilibrium distribution function in the following way \cite{sedrakianPRD},
 
 \begin{align}
  \delta f = (\vec{p}.~\vec{\Xi})\frac{\partial f_0}{\partial\epsilon},
  \label{equ24}
 \end{align}
with
\begin{align}
 \vec{\Xi}= a \vec{e}+ b\vec{h} +c (\vec{e}\times \vec{h}),
 \label{equ25}
\end{align}
where, $\vec{e}=\frac{\vec{E}}{|E|}$ and $\vec{h}=\frac{\vec{B}}{|B|}$, are the direction of the electric field and magnetic field respectively. This is similar to the ansatz taken as in Eq.\eqref{equ5} for the calculation of thermal conductivity, with $\vec{\nabla}T$ now being replaced by $\vec{e}$. Using Eq.\eqref{equ24} and \eqref{equ25}, Eq.\eqref{equ23} can be expressed as,
\begin{align}
 q(\vec{E}.\vec{v})-qBa\vec{v}.(\vec{e}\times\vec{h})-qBc(\vec{e}.\vec{h})(\vec{v}.\vec{h})+qBc(\vec{v}.\vec{e})= -\frac{\epsilon}{\tau}\bigg[a (\vec{v}.\vec{e})+b(\vec{v}.\vec{h})+c \vec{v}.(\vec{e}\times\vec{h})\bigg]
 \label{equ26}
\end{align}

Comparing coefficients of difference tensor structures in Eq.\eqref{equ26} we get,
\begin{align}
 c= \frac{qB}{\epsilon}\tau a \equiv \omega_c\tau a,
 \label{equ27}
\end{align}
\begin{align}
b= (\omega_c\tau)^2 a (\vec{e}.\vec{h}), 
\label{equ28}
\end{align}

and,
\begin{align}
 qBc + qE=-\frac{\epsilon a}{\tau}.
 \label{equ29}
\end{align}
Using Eq.\eqref{equ27}, Eq.\eqref{equ28} and Eq.\eqref{equ29} it can be shown that,

\begin{align}
 a =\frac{-qE}{1+(\omega_c\tau)^2}\bigg(\frac{\tau}{\epsilon}\bigg).
 \label{equ30}
\end{align}

Hence,
\begin{align}
 \delta f = -\frac{q \tau}{1+(\omega_c\tau)^2}\bigg[\vec{v}.\vec{E}+(\omega_c\tau)\vec{v}.(\vec{E}\times\vec{h})+(\omega_c\tau)^2(\vec{v}.\vec{h})(\vec{E}.\vec{h})\bigg]\frac{\partial f_0}{\partial\epsilon}.
\label{equ31}
 \end{align}

The electric current ($\vec{j}$) can be defined as,
\begin{align}
 \vec{j} = \int \frac{d^3p}{(2\pi)^3}~q~\vec{v}~\delta f
\label{equ32}
 \end{align}

 Using Eq.\eqref{equ31}, electric current as given in Eq.\eqref{equ32} can be expressed as,
 \begin{align}
  j^l  & = \frac{q^2}{3}\int\frac{d^3p}{(2\pi)^3} \frac{v^2\tau}{1+(\omega_c\tau)^2}\bigg[E^l+(\omega_c\tau)\epsilon^{ljk}h^kE^j+(\omega_c\tau)^2h^lh^jE^j\bigg]\bigg(-\frac{\partial f_0}{\partial\epsilon}\bigg)\nonumber\\
  & = \bigg(\sigma_0\delta^{lj}-\sigma_1\epsilon^{lkj}h^k+\sigma_2h^lh^j\bigg)E^j.
  \label{equ33}
   \end{align}
   From Eq.\eqref{equ33} we can identify various components of electrical conductivity tensor in the presence of magnetic field,   
   \begin{align}
    \sigma_0 = \frac{q^2}{3T}\int \frac{d^3p}{(2\pi)^3}\tau \bigg(\frac{p^2}{\epsilon^2}\bigg)\frac{1}{1+(\omega_c\tau)^2}f_0,
    \label{equ34}
   \end{align}
   \begin{align}
    \sigma_1 = \frac{q^2}{3T}\int \frac{d^3p}{(2\pi)^3}\tau \bigg(\frac{p^2}{\epsilon^2}\bigg)\frac{\omega_c\tau}{1+(\omega_c\tau)^2}f_0,
    \label{equ35}
   \end{align}
   \begin{align}
    \sigma_2 = \frac{q^2}{3T}\int \frac{d^3p}{(2\pi)^3}\tau \bigg(\frac{p^2}{\epsilon^2}\bigg)\frac{(\omega_c\tau)^2}{1+(\omega_c\tau)^2}f_0.
    \label{equ36}
   \end{align}
   
It is important to note that in Ref.\cite{hallhrg,hallqgp} we discussed electrical conductivity and Hall conductivity in the presence of magnetic field.
$\sigma_0$ and $\sigma_1$ as given in Eq.\eqref{equ34} and \eqref{equ35} can be identified with electrical conductivity in the presence of magnetic field and Hall conductivity respectively as obtain in Ref\cite{hallhrg}. 
In Ref.\cite{hallhrg,hallqgp} we only considered electric field and magnetic field perpendicular to each other, 
however for a general configuration of electric and magnetic field we get another transport coefficient $\sigma_2$ \cite{sedrakianPRD}. Hence for a general configurations of electric and magnetic field we have three different components of electrical conductivity tensor. For a multi component model 
total $\sigma_0$, $\sigma_1$ and $\sigma_2$ can be expressed as,
  \begin{align}
    \sigma_0 = \sum_i\frac{q_i^2g_i}{3T}\int \frac{d^3p}{(2\pi)^3} \bigg(\frac{p^2}{\epsilon_i^2}\bigg)\frac{1}{1+(\omega_{c_i}\tau_i)^2}f_{0_i}\tau_i,
    \label{equ37}
   \end{align}
   \begin{align}
    \sigma_1 = \sum_i\frac{q_i^2g_i}{3T}\int \frac{d^3p}{(2\pi)^3} \bigg(\frac{p^2}{\epsilon_i^2}\bigg)\frac{\omega_{c_i}\tau_i}{1+(\omega_{c_i}\tau_i)^2}f_{0_i}\tau_i,
    \label{equ38}
   \end{align}
   \begin{align}
    \sigma_2 = \sum_i\frac{q_i^2g_i}{3T}\int \frac{d^3p}{(2\pi)^3} \bigg(\frac{p^2}{\epsilon_i^2}\bigg)\frac{(\omega_{c_i}\tau_i)^2}{1+(\omega_{c_i}\tau_i)^2}f_{0_i}\tau_i.
    \label{equ39}
   \end{align}
Here $\epsilon_i$, $g_i$, $\tau_i$ are the single particle dispersion relation, degeneracy factor and relaxation time of ``i-th'' particle species. 
From Eq.\eqref{equ38} it is clear that Hall conductivity is zero at vanishing baryon chemical potential even at finite magnetic field. Only at finite baryon chemical potential Hall conductivity has non vanishing value at finite magnetic field. Behaviour of $\sigma_0$ and $\sigma_1$ with temperature, baryon chemical potential and magnetic field has been discussed in Ref.\cite{hallhrg}. In this investigation we present variation of $\sigma_2$ with temperature, baryon chemical potential and magnetic field.

\section{Shear viscosity in the presence of magnetic field}
\label{shearvisco}
Effect of magnetic field on shear viscosity of strongly interacting matter has discussed in Ref.\cite{landaubook,tuchin3,sabya1,sabya2,sabya3,payal}. Without going into the details of the formalism, for completeness we briefly mention here the salient features of the formalism. Following Ref.\cite{landaubook,tuchin3,sabya1,sabya2,sabya3,payal}we start with Boltzmann kinetic equation in the presence of magnetic field as discussed in Ref.\cite{tuchin3},

\begin{align}
 p^{\mu}\partial_{\mu}f_0+qB^{\mu\nu}\frac{\partial\delta f}{\partial u^{\mu}}u_{\nu}=\mathcal{C}[\delta f],
 \label{equ40new}
\end{align}
here $f_0$ is the equilibrium distribution function. In the Boltzmann approximation, $f_0=e^{-p^{\mu}U(x)_{\mu}/T\pm\mu_B/T}$. $U^\lambda \equiv (\gamma_V,\gamma_V\vec{V})$ is the macroscopic velocity of the fluid, $p^{\mu}=mu^{\mu}$ is the particle four momentum. $\delta f$ is the deviation from equilibrium, $B^{\mu\nu}$ is electromagnetic 
field tensor which contain magnetic field \cite{tuchin3}. In Boltzmann approximation, 
\begin{align}
 \partial_{\mu}f_0 = -\frac{f_0}{T}p^{\lambda}\partial_{\mu}U_{\lambda}(x).
\end{align}
In comoving frame, it can be shown that\cite{tuchin3},
\begin{align}
 \partial_{\mu}f_0|_{\vec{V}=0} = - \frac{f_0}{T}p_{\nu}\partial_{\mu}V^{\nu}.
 \label{equ42new}
\end{align}
Using Eq.\eqref{equ42new}, Boltzmann equation \eqref{equ40new} can be written as,
\begin{align}
 -\frac{f_0}{T}p^{\mu}p^{\nu}V_{\mu\nu} =- qB^{\mu\nu}\frac{\partial \delta f}{\partial u^{\mu}}u_{\nu}+\mathcal{C}[\delta f],
\end{align}
here, $V^{\mu\nu}=\frac{1}{2}(\partial^{\mu}V^{\nu}+\partial^{\nu}V^{\mu})$. Shear viscosity only deals with spatial variation of fluid velocity. Hence we can drop all temporal dependence from the Boltzmann equation and only deal with spatial derivatives. Hence the Boltzmann equation considering only the spatial derivatives of fluid velocity is, 

\begin{align}
 \frac{\epsilon}{T}p_{\alpha}v_{\beta}V_{\alpha\beta}f_0 = \frac{qB}{\epsilon} b_{\alpha\beta}v_{\beta}\frac{\partial(\delta f)}{\partial v^{\alpha}}
 +\frac{\delta f}{\tau},
 \label{equ40}
\end{align}
here $\alpha,\beta$ are spatial indices, $b_{\alpha\beta}\equiv\epsilon_{\alpha\beta\gamma}b_{\gamma}$, $b_{\gamma}=\frac{B_{\gamma}}{B}$, $B^{\mu\nu}$ is the electromagnetic field tensor with $\mu,\nu=\{0,1,2,3\}$. $V_{\alpha\beta}=\frac{1}{2}(\partial_{\alpha}V_{\beta}+\partial_{\beta}V_{\alpha})$, $V_{\alpha}$ is the fluid velocity, $p^{\mu}=(\epsilon,\vec{p})$ is the particle four momentum. Following Ref.\cite{landaubook,tuchin3} we can express $\delta f$ in the following manner,
\begin{align}
 \delta f = \sum_{n=0}^{n=4} g_{(n)}V^{(n)}_{\alpha\beta}v_{\alpha}v_{\beta},
 \label{equ41}
\end{align}
where the tensors $V^{(n)}_{\alpha\beta}$ are,
\begin{align}
& V^{(0)}_{\alpha\beta}=(3b_{\alpha}b_{\beta}-\delta_{\alpha\beta})(b_{\gamma}b_{\delta}V_{\gamma\delta}-\frac{1}{3}\vec{\nabla}.\vec{V}),\nonumber\\
& V^{(1)}_{\alpha\beta} = 2V_{\alpha\beta}+\delta_{\alpha\beta}V_{\gamma\delta}b_{\gamma}b_{\delta}-2V_{\alpha\gamma}b_{\gamma}b_{\beta}-2V_{\beta\gamma}b_{\gamma}b_{\alpha}+(b_{\alpha}b_{\beta}-\delta_{\alpha\beta})\vec{\nabla}.\vec{V}+b_{\alpha}b_{\beta}V_{\gamma\delta}b_{\gamma}b_{\delta},\nonumber\\
&V^{(2)}_{\alpha\beta}= 2(V_{\alpha\gamma}b_{\beta}b_{\gamma}+V_{\beta\gamma}b_{\alpha}b_{\gamma}-2b_{\alpha}b_{\beta}V_{\gamma\delta}b_{\gamma}b_{\delta}),\nonumber\\
&V^{(3)}_{\alpha\beta}= V_{\alpha\gamma}b_{\beta\gamma}+V_{\beta\gamma}b_{\alpha\gamma}-V_{\gamma\delta}b_{\alpha\gamma}b_{\beta}b_{\delta}-V_{\gamma\delta}b_{\beta\gamma}b_{\alpha}b_{\delta},\nonumber\\
&V^{(4)}_{\alpha\beta}= 2(V_{\gamma\delta}b_{\alpha\gamma}b_{\beta}b_{\delta}+V_{\gamma\delta}b_{\beta\gamma}b_{\alpha}b_{\delta}). 
\label{equ42}
\end{align}

Deviation of purely spatial components of the  energy momentum tensor from equilibrium energy momentum tensor can be written as \cite{tuchin3,sabya1},
\begin{align}
 \delta T_{\alpha\beta}=\int\frac{d^3p}{(2\pi)^3}v_{\alpha}v_{\beta}\epsilon\delta f.
 \label{equ49}
\end{align}
Again using the functions $V^{(n)}_{\alpha\beta}$ we can write,
\begin{align}
 \delta T_{\alpha\beta}=\sum_{n=0}^4 \eta_{n}V^{(n)}_{\alpha\beta}
 \label{equ50}
\end{align}

The viscosity component associated with the tensor $V_{\alpha\beta}^{(0)}$ is the longitudinal viscosity as $V_{\alpha\beta}^{(0)}b_{\alpha}b_{\beta}\neq0$ while the components $\eta^{(n)}$ corresponding to $V_{\alpha\beta}^{(n)}$ $(n=1,2,3,4)$ are called the transverse viscosities as they are transverse to $b_{\alpha}b_{\beta}$.
To calculate transverse shear viscosity
coefficients we impose the conditions, $\vec{\nabla}.\vec{V}=0$ and $V_{\gamma\delta}b_{\gamma}b_{\delta} = 0$ \cite{landaubook,tuchin3}. Hence $V^{(0)}_{\alpha\beta} = 0$. Using 
the tensors $V^{(n)}_{\alpha\beta}$, $n=1,2,3,4$, Eq.\eqref{equ40} can be expressed as (for details see Ref.\cite{tuchin3,sabya1,sabya2,sabya3}),
\begin{align}
 \frac{\epsilon}{T}v_{\alpha}v_{\beta}V_{\alpha\beta}f_0 & = 2\omega_c g_1\bigg[2V_{\alpha\gamma}b_{\alpha\beta}v_{\beta}v_{\gamma}-2V_{\alpha\rho}b_{\alpha\beta}b_{\rho}v_{\beta}(\vec{b}.\vec{v})\bigg]+2\omega_c g_2\bigg[2V_{\alpha\rho} b_{\alpha\beta}v_{\beta}b_{\rho}(\vec{v}.\vec{b})\bigg]\nonumber\\
 & +2\omega_c g_3\bigg[2V_{\alpha\beta}v_{\alpha}v_{\beta}-4V_{\alpha\beta}v_{\alpha}b_{\beta}(\vec{b}.\vec{v})\bigg]
 +2\omega_c g_4 \bigg[2V_{\alpha\beta}v_{\alpha}b_{\beta}(\vec{b}.\vec{v})\bigg]\nonumber\\
 & +\frac{g_1}{\tau}\bigg[2V_{\gamma\delta}v_{\gamma}v_{\delta}-4V_{\gamma\rho}v_{\gamma}b_{\rho}(\vec{b}.\vec{v})\bigg]+\frac{g_2}{\tau}\bigg[4V_{\gamma\rho}v_{\gamma}b_{\rho}(\vec{b}.\vec{v})\bigg]\nonumber\\
 & +\frac{g_3}{\tau}\bigg[2V_{\gamma\rho}b_{\delta\rho}v_{\gamma}v_{\delta} -2 V_{\rho\sigma}b_{\gamma\rho}b_{\sigma}v_{\gamma}(\vec{b}.\vec{v})\bigg]+\frac{g_4}{\tau}\bigg[4 V_{\rho\sigma}b_{\gamma\rho}b_{\sigma}v_{\gamma}(\vec{b}.\vec{v})\bigg]
 \label{equ43}
\end{align}
To get Eq.\eqref{equ43} we have used $\vec{\nabla}.\vec{V}=0$, $V_{\alpha\beta}b_{\alpha}b_{\beta}=0, b_{\alpha\beta}b_{\alpha}=0, b_{\alpha\beta}v_{\alpha}v_{\beta}=0$ and $b_{\alpha}b_{\alpha}=1$.
Comparing various tensor structure in Eq.\eqref{equ43} we can write,
\begin{align}
 & g_3=2\omega_c\tau g_1\nonumber\\
 & 2\omega_c g_3-\omega_c g_4+\frac{g_1}{\tau}-\frac{g_2}{\tau} = 0\nonumber\\
 & 4\omega_c g_1 -4\omega_cg_2+\frac{2}{\tau}g_3-\frac{4}{\tau}g_4=0\nonumber\\
 & 2\omega_cg_3+\frac{1}{\tau}g_1=\frac{\epsilon}{2T}f_0.
\label{equ44}
 \end{align}
Solving above set of equations for the coefficients
$g_1,g_2,g_3$ and $g_4$ we get,
\begin{align}
 g_1=\frac{\epsilon}{2T}\frac{\tau}{\bigg[1+4(\omega_c\tau)^2\bigg]}f_0,
 \label{equ45}
\end{align}
\begin{align}
g_2=\frac{\epsilon}{2T}\frac{\tau}{\bigg[1+(\omega_c\tau)^2\bigg]}f_0,
\label{equ46}
\end{align}
\begin{align}
 g_3=\frac{\epsilon}{2T}\frac{\tau(\omega_c\tau)}{\bigg[\frac{1}{2}+2(\omega_c\tau)^2\bigg]}f_0,
 \label{equ47}
\end{align}
\begin{align}
 g_4=\frac{\epsilon}{2T}\frac{\tau(\omega_c\tau)}{\bigg[1+(\omega_c\tau)^2\bigg]}f_0,
\label{equ48}
 \end{align}

Using Eq.\eqref{equ49} and Eq.\eqref{equ50} various components of the shear viscosity in magnetic field can be shown to be \cite{tuchin3,sabya1},
\begin{align}
 \eta_n=\frac{2}{15}\int\frac{d^3p}{(2\pi)^3}\epsilon g_{n}v^4, ~~~n=1,2,3,4
 \label{equ51}
\end{align}
hence
\begin{align}
 \eta_1= \frac{1}{15T}\int\frac{d^3p}{(2\pi)^3}\frac{p^4}{\epsilon^2} \frac{\tau}{1+4(\omega_c\tau)^2}f_0
\label{equ52}
 \end{align}
\begin{align}
 \eta_2 = \frac{1}{15T}\int\frac{d^3p}{(2\pi)^3}\frac{p^4}{\epsilon^2} \frac{\tau}{1+(\omega_c\tau)^2}f_0
 \label{equ53}
\end{align}
\begin{align}
 \eta_3 = \frac{1}{15T}\int\frac{d^3p}{(2\pi)^3}\frac{p^4}{\epsilon^2} \frac{\tau(\omega_c\tau)}{\frac{1}{2}+2(\omega_c\tau)^2}f_0
 \label{equ54}
\end{align}
\begin{align}
 \eta_4 = \frac{1}{15T}\int\frac{d^3p}{(2\pi)^3}\frac{p^4}{\epsilon^2} \frac{\tau(\omega_c\tau)}{1+(\omega_c\tau)^2}f_0
 \label{equ55}
\end{align}

For hadron resonance gas model total shear viscosity in the presence of magnetic field can be expressed as, 
\begin{align}
 \eta_1= \sum_i\frac{g_i}{15T}\int\frac{d^3p}{(2\pi)^3}\frac{p^4}{\epsilon_i^2} \frac{1}{1+4(\omega_{c_i}\tau_i)^2}f_{0_i}\tau_i,
 \label{equ56}
\end{align}
\begin{align}
 \eta_2 = \sum_i\frac{g_i}{15T}\int\frac{d^3p}{(2\pi)^3}\frac{p^4}{\epsilon_i^2} \frac{1}{1+(\omega_{c_i}\tau_i)^2}f_{0_i}\tau_i,
 \label{equ57}
\end{align}
\begin{align}
 \eta_3 = \sum_i\frac{g_i}{15T}\int\frac{d^3p}{(2\pi)^3}\frac{p^4}{\epsilon_i^2} \frac{(\omega_{c_i}\tau_i)}{\frac{1}{2}+2(\omega_{c_i}\tau_i)^2}f_{0_i}\tau_i,
 \label{equ58}
\end{align}
\begin{align}
 \eta_4 = \sum_i\frac{g_i}{15T}\int\frac{d^3p}{(2\pi)^3}\frac{p^4}{\epsilon_i^2} \frac{(\omega_{c_i}\tau_i)}{1+(\omega_{c_i}\tau_i)^2}f_{0_i}\tau_i.
 \label{equ59}
\end{align}
In Eq.\eqref{equ56},\eqref{equ57},\eqref{equ58},\eqref{equ59} $g_i$, $\tau_i$ are degeneracy factor and relaxation time of ``i-th'' hadron. $g_i$ in Eq.\eqref{equ56},\eqref{equ57},\eqref{equ58},\eqref{equ59} should not be confused with $g_1, g_2,g_3, g_4$ as given is Eq.\eqref{equ45},\eqref{equ46},\eqref{equ47}, \eqref{equ48}.
In the absence of magnetic field only $V_{\alpha\beta}^{(1)}$ is non vanishing and the corresponding shear viscosity is $\eta_1$. In the absence of magnetic field $\eta_1=\eta_2$ and $\eta_3=\eta_4=0$. $\eta_3$ and $\eta_4$ are Hall type shear viscosities. Similar
to other Hall type transport coefficients, Hall type shear viscosity also vanishes for zero baryon chemical potential even for non vanishing magnetic field. This is because, at vanishing baryon chemical potential, when the number density of particles and anti particles are same,  particles and antiparticles will have equal and opposite contribution to shear viscosity $\eta_3$ and $\eta_4$. Only at finite baryon chemical potential $\eta_3$ and $\eta_4$ can take non vanishing values.
In this context a comment regarding the anisotropic $\eta$ may be in order. Firstly let us note that $\eta_i (i=1,2,3,4)$ is smaller compared to the longitudinal viscosity coefficient $\eta_{0}$. Therefore, the flow velocity in the direction perpendicular to the direction of the magnetic filed will be larger as compared to the case in the absence of magnetic field.

From Eq.\eqref{equ19}-Eq.\eqref{equ21}, Eq.\eqref{equ37}-Eq.\eqref{equ39} and Eq.\eqref{equ56}-Eq.\eqref{equ59}; it is clear that the important input required for the estimation of the transport coefficient is the relaxation time $\tau^i$ which is in general can be energy dependent. In this investigation we consider only energy averaged relaxation time. Further, the coefficients of thermal conductivity are dependent on bulk thermodynamic properties of the system e.g. energy density, pressure and enthalpy. These thermodynamic quantities and the relaxation time will be estimated for the hadronic system within the hadron resonance gas model that we discuss in the next section.

\section{HADRON RESONANCE GAS MODEL}
\label{HRGmodel}
The thermodynamic potential of a non interacting gas of hadrons and its resonances at finite temperature ($T$) and baryon chemical potential ($\mu_B$) can be expressed as \cite{KadamHM2015},

\begin{equation}
 \log Z(\beta,\mu_B,V)=\int dm (\rho_M(m)\log Z_b(m,V,\beta,\mu_B)+\rho_B(m)\log Z_f(m,V,\beta,\mu_B)),
 \label{equ60}
\end{equation}
where, $V$ is the volume and $T = 1/\beta$ is the temperature of point like hadrons and their resonances. Total partition function is sum of the partition functions of free mesons ($Z_b$) and baryons ($Z_f$) with mass $m$.  
Moreover, spectral function which encodes hadron properties are represented as $\rho_B$ and $\rho_M$ for free mesons and baryons respectively. Various thermodynamic quantities can be calculated using derivatives of the 
logarithm of the partition function as given in 
Eq.\eqref{equ60}, with respect to the thermodynamic parameters T, $\mu_B$ and the volume $V$. In this investigation we confine ourselves to ideal HRG model where we have considered all the hadrons and their resonances below a certain mass cutoff $\Lambda$. This can be achieved by taking the following form of spectral density, 

\begin{equation}
 \rho_{B/M}(m)=\sum_i^{m_i<\Lambda}g_i\delta(m-m_i),
 \label{equ61}
\end{equation}
here $m_i$ and $g_i$ are mass and degeneracy of ``i-th'' hadron species. Although HRG with discrete particle spectrum is very appealing because of its simple structure, but
it can  explain
lattice QCD data for trace anomaly only up to temperature $\sim 130 $ MeV \cite{HRGMuller}. Including Hagedron spectrum along with discrete particle spectrum HRG model can explain lattice QCD data for QCD trace anomaly up to $T \sim$ 160 MeV \cite{HRGMuller}. For details of thermodynamics of HRG model, see e.g. Ref.\cite{HRG1}. 

The relaxation time of particle $a$ with three momentum $p_a$ and energy $\epsilon_a$ is expressed as \cite{lataguruhm,KapustaChakraborty2011},

\begin{align}
\tau_a^{-1}(\epsilon_a)=\sum_{b,c,d}\int\frac{d^3p_b}{(2\pi)^3}\frac{d^3p_c}{(2\pi)^3}\frac{d^3p_d}{(2\pi)^3}W(a,b\rightarrow c,d)f_b^0,
\label{equ62}
\end{align}
here $W(a,b\rightarrow c,d)$ is the transition rate for process $a(p_a)+b(p_b)\rightarrow c(p_c)+d(p_d)$, can be expressed in terms of the transition amplitude $\mathcal{M}$ in the following way,

\begin{align}
W(a,b\rightarrow c,d)=\frac{(2\pi)^4\delta(p_a+p_b-p_c-p_d)}{2\epsilon_a2\epsilon_b2\epsilon_c2\epsilon_d}|\mathcal{M}|^2.
\label{equ63}
\end{align}

In the center of mass frame, the relaxation time ($\tau_a$) or equivalently interaction frequency ($\omega_a$) can be written as, 

\begin{align}
\tau_a^{-1}(\epsilon_a)\equiv \omega_a(\epsilon_a)=\sum_b\int\frac{d^3p_b}{(2\pi)^3}\sigma_{ab}v_{ab}f_b^0.
\label{equ64}
\end{align}

Here $\sigma_{ab}$ is the total scattering cross section for the process, $a(p_a)+b(p_b)\rightarrow c(p_c)+d(p_d)$ and $v_{ab}$ is the relativistic relative velocity between particle $a$ and $b$,

\begin{align}
 v_{ab}=\frac{\sqrt{(p_a.p_b)^2-m_a^2m_b^2}}{\epsilon_a\epsilon_b}.
 \label{equ65}
\end{align}

In this work we shall be considering energy averaged relaxation time. One can obtain the energy independent relaxation time $\tau^a$ by averaging the relaxation time $\tau^a(\epsilon_a)$ over the distribution function $f_a^0(\epsilon_a)$ \cite{KapustaChakraborty2011,paramitahm2016},

\begin{align}
 \tau_a^{~-1}=\frac{\int \frac{d^3p_a}{(2\pi)^3} f_a^0\tau^{-1}_a(\epsilon_a)}{\int \frac{d^3p_a}{(2\pi)^3} f^0_a}=\sum_b \frac{\int \frac{d^3p_a}{(2\pi)^3} \frac{d^3p_b}{(2\pi)^3}  f_a^0f_b^0 \sigma_{ab}v_{ab}}{\int \frac{d^3p_a}{(2\pi)^3} f^0_a}
 \label{equ66}
\end{align}

Using Eq.\eqref{equ66}, the energy averaged relaxation time
$(\tau_a)$,  can be expressed as\cite{GURUHM2015},

\begin{equation}
 \tau_a^{~-1}=\sum_b n_b\langle\sigma_{ab}v_{ab}\rangle, ~~~\text{where},~~n_b=\int\frac{d^3p_b}{(2\pi)^3}f_b^0
 \label{equ67}
\end{equation}
here $n_b$ and $\langle\sigma_{ab}v_{ab}\rangle$ represents number density and thermal averaged cross section respectively. The thermal averaged 
cross section for the scattering process $a(p_a)+b(p_b)\rightarrow c(p_c)+d(p_d)$ is given as \cite{gondologelmini},

\begin{align}
 \langle\sigma_{ab}v_{ab}\rangle = \frac{\int d^3p_ad^3p_b \sigma_{ab}v_{ab}f_a^{0}(p_a)f_a^{0}(p_b)}{\int d^3p_ad^3p_b f_a^{0}(p_a)f_a^{0}(p_b)}.
 \label{equ68}
\end{align}
In Boltzmann approximation and for hard sphere (of radius $r_h$) scattering for the cross section  ($\sigma=4\pi r_h^2$) the thermal averaged cross section becomes, 
\begin{align}
 \langle\sigma_{ab}v_{ab}\rangle = \frac{\sigma\int d^3p_ad^3p_b v_{ab}e^{-\epsilon_a/T}e^{-\epsilon_b/T}}{\int d^3p_ad^3p_b e^{-\epsilon_a/T}e^{-\epsilon_b/T}}.
 \label{equ69}
\end{align}

Note that in Boltzmann approximation of the thermal averaged relaxation time chemical potential dependence gets canceled from the numerator and denominator. Rather than using  momentum integration it is useful to introduced center of mass energy variable
($\sqrt{s}$) to calculate thermal averaged cross section. In terms of center of mass energy variable ($\sqrt{s}$) it can be shown that, 

\begin{align}
 \int d^3p_ad^3p_b v_{ab}e^{-\epsilon_a/T}e^{-\epsilon_b/T} = 2\pi^2T\int ds\sqrt{s}(s-4m^2)K_1(\sqrt{s}/T),
 \label{equ70}
\end{align}
and
\begin{align}
 \int d^3p_ad^3p_b e^{-\epsilon_a/T}e^{-\epsilon_b/T} = \left(4\pi m^2 T K_2(m/T)\right)^2.
 \label{equ71}
\end{align}

Thus the thermal averaged cross section can be given as, 

\begin{align}
 \langle\sigma_{ab}v_{ab}\rangle =\frac{\sigma}{8m^4TK_2^2(m/T)}\int_{4m^2}^{\infty}ds \sqrt{s}(s-4m^2)K_1(\sqrt s /T).
\label{equ72}
 \end{align}

Here $\sqrt{s}$ is the center of mass energy, $K_1$ and $K_2$ are modified Bessel function of first order and second order respectively. When the
particles are of different species then the above equation can be generalized as, 

\begin{equation}
 \langle\sigma_{ab}v_{ab}\rangle = \frac{\sigma}{8Tm_a^2m_b^2K_2(m_a/T)K_2(m_b/T)}\int_{(m_a+m_b)^2}^{\infty}ds\times \frac{[s-(m_a-m_b)^2]}{\sqrt{s}}
 \times [s-(m_a+m_b)^2]K_1(\sqrt{s}/T),
 \label{equ73}
\end{equation}
where $\sigma = 4\pi r_h^2$ is the total scattering cross section for the hard sphere. It is important to mention that in hard sphere scattering approximation section $\sigma$ is independent of temperature and baryon chemical potential. But thermal averaged cross section $\langle\sigma v\rangle$
 is in general can depend on temperature ($T$) and chemical potential $\mu_B$. Only in  Boltzmann approximation $\langle\sigma v\rangle$ is independent of $\mu_B$  \cite{gondologelmini}. Evaluating the thermal averaged relaxation time for each species we 
estimate various transport coefficients of the hot and dense hadron gas.

\section{results and discussions}
\label{results}
We have estimated thermal
conductivity,electrical conductivity and shear viscosity in the presence of magnetic field within the framework of hadron resonance gas model.
For the hadron resonance gas model, we consider all the hadrons and their resonances up to a mass cutoff $\Lambda$ which we take as $\Lambda =  2.6$ GeV as is listed in Ref.\cite{pdg}. For a detailed list of hadrons and its resonances we refer to  Appendix A of Ref.\cite{kapustaAlbr}. This apart radii of the hard spheres also enters in the calculation of  relaxation time. We have considered an uniform radius of $r_h=0.5$ fm for all the hadrons
\cite{GURUHM2015,hmgururadius1}. With these set of parameters, we have estimated thermal conductivity, shear viscosity etc. as a function of temperature $(T)$ and baryon chemical potential ($\mu_B$) for different values of the magnetic field $(B)$.

\subsection{Results for thermal conductivities in a magnetic field}

\begin{figure}[!htp]
\begin{center}
\includegraphics[width=0.6\textwidth]{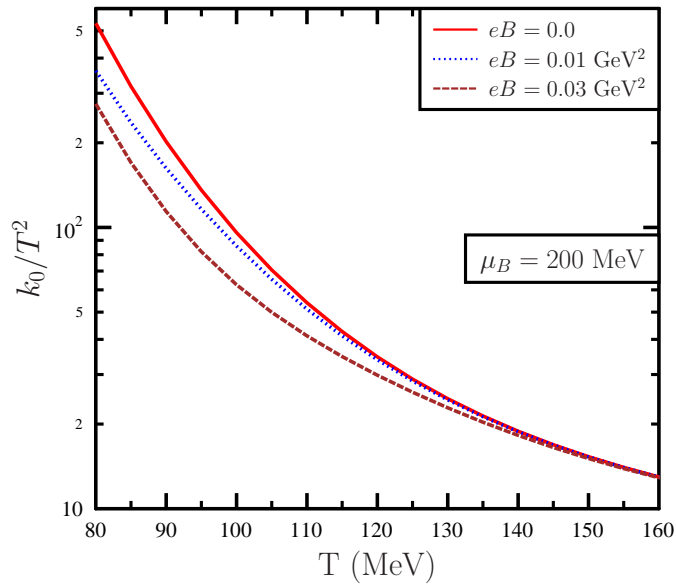}
\caption{Variation of normalized thermal conductivity ($k_0/T^2$) with temperature ($T$) for different values of magnetic field ($B$) at finite baryon chemical potential. Red solid line represents $B=0$ case, blue dotted line and brown dashed line represents $eB=0.01$ GeV$^2$ and $eB=0.03$ GeV$^2$ respectively. In the presence of magnetic field $k_0/T^2$ decreases. At low temperature decrease in $k_0/T^2$ due to magnetic field is significant. But at higher temperature effect of magnetic field on $k_0/T^2$ is not significant.}
\label{fig1}
\end{center}
\end{figure}

\begin{figure}[!htp]
\begin{center}
\includegraphics[width=0.49\textwidth]{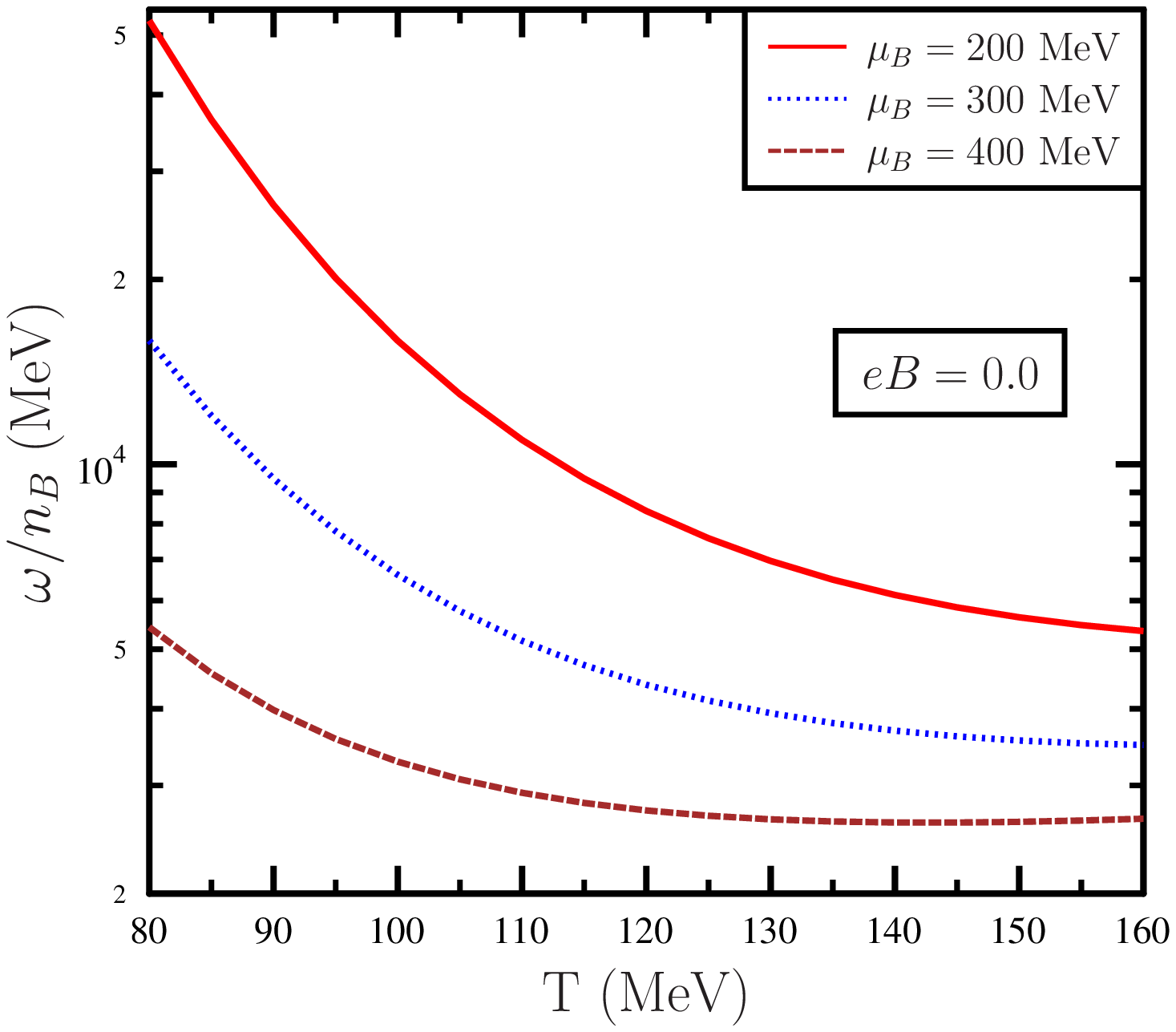}
\includegraphics[width=0.49\textwidth]{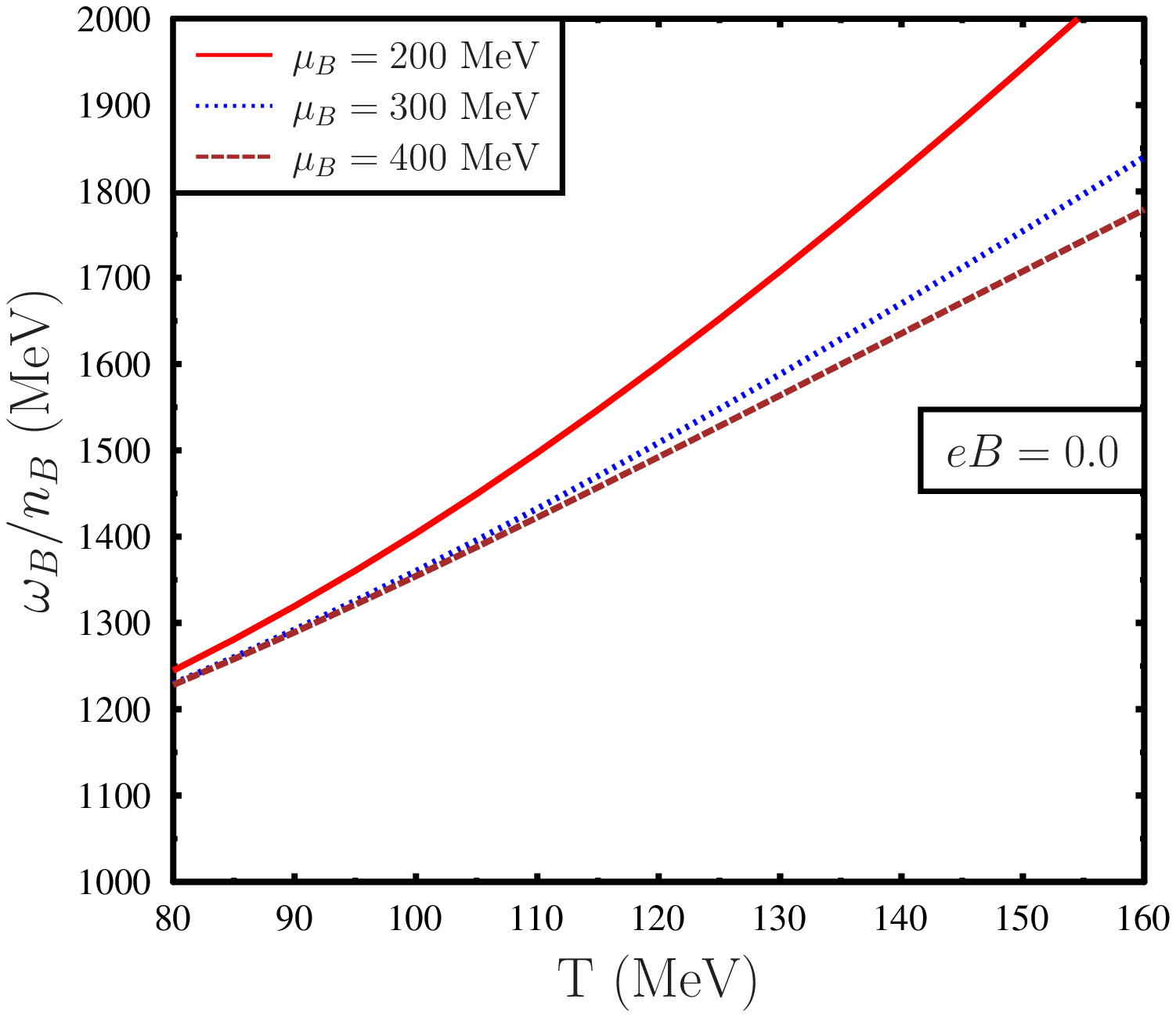}
\caption{Left plot: Variation of $\omega/n_B$ with temperature ($T$) for different values of $\mu_B$  at zero magnetic field. Right plot:Variation of $\omega/n_B$ only for baryons denoted as $\omega_B/n_B$, with temperature ($T$) for different values of $\mu_B$ for a vanishing magnetic field.  With increasing temperature and $\mu_B$, $\omega/n_B$ of the hadron resonance gas decreases. However baryonic contribution to $\omega/n_B$ i.e. $\omega_B/n_B$ increases with temperature. }
\label{fig2new}
\end{center}
\end{figure}

In Fig.\eqref{fig1} we show the variation of normalized thermal conductivity ($k_0/T^2$) with temperature $(T)$ for various values of magnetic field at a finite baryon chemical potential. As may be observed from the figure the normalized thermal conductivity $k_0/T^2$ decreases with temperature. Let us note that $k_0/T^2$ as given in Eq.\eqref{equ19} depends on the relaxation time, $\omega/n_B$ and distribution function. As temperature increases the scattering rate increases as the number of particle increases. This leads to relaxation time, which is inverse of scattering rate, decreasing with temperature. Further $\omega/n_B$ also decreases with temperature which has been shown in the left plot in Fig.\eqref{fig2new}. The reason for this behavior of $\omega/n_B$ with temperature can be understood as follows. The dominant contribution to the sum over all hadrons arise from pions and protons which can be approximately given by,
\begin{align}
 \frac{\omega}{n_B}=\frac{\mathcal{E}+P}{n_B}\sim\frac{\mathcal{E}_{\pi}+P_{\pi}}{n_p}+\frac{\mathcal{E}_{p}+P_{p}}{n_p}\sim \frac{e^{-m_{\pi}/T}(m_{\pi}+T)}{\sinh(\mu_B/T)e^{-m_{p}/T}}
+\frac{\cosh(\mu_B/T)e^{-m_{p}/T}(m_{p}+T)}{\sinh(\mu_B/T)e^{-m_{p}/T}}.
\label{equ74}
 \end{align}
With increasing temperature $\coth(\mu_B/T)$ as well as $(m_{p}+T)$ increases. Hence if one considers only baryons then with temperature $\omega_{B}/n_B$ increases as can be seen in the right plot of Fig.\eqref{fig2new}. However for pions $\omega_{\pi}/n_B$ decreases with temperature due to the term $e^{(m_p-m_{\pi})/T}$ in Eq.\eqref{equ74}. For hadron resonance gas contributions of mesons in the energy density and pressure is significantly large with respect to the baryonic contributions. Hence when we consider hadron resonance gas, due to mesonic contribution to energy density and pressure, $\omega/n_B$ decreases with temperature as can be seen in the left plot in Fig.\eqref{fig2new}.

  It is also clear that in the presence of magnetic field thermal conductivity decreases. This can be understood from the expression for $k_0/T^2$ in Eq.\eqref{equ19} which is inversely proportional to $1+(\omega_c\tau)^2$.
     At low temperature relaxation time is  relatively larger and at low temperature $k_0\sim \frac{1}{\omega_c^2\tau}$. Hence at low temperature magnetic affects $k_0/T^2$ significantly. On the other hand at high temperature $\tau$ is small hence effect of $\omega_c\tau$ in the denominator of Eq.\eqref{equ19} is not significant. Thus at small temperature  $k_0/T^2$ decreases with magnetic field but at large temperature magnetic field does not affect $k_0/T^2$ significantly. This behaviour of thermal conductivity is analogous to the variation of electrical conductivity ($\sigma_0/T$) as discussed in Ref.\cite{hallhrg}.  

\begin{figure}[!htp]
\begin{center}
\includegraphics[width=0.49\textwidth]{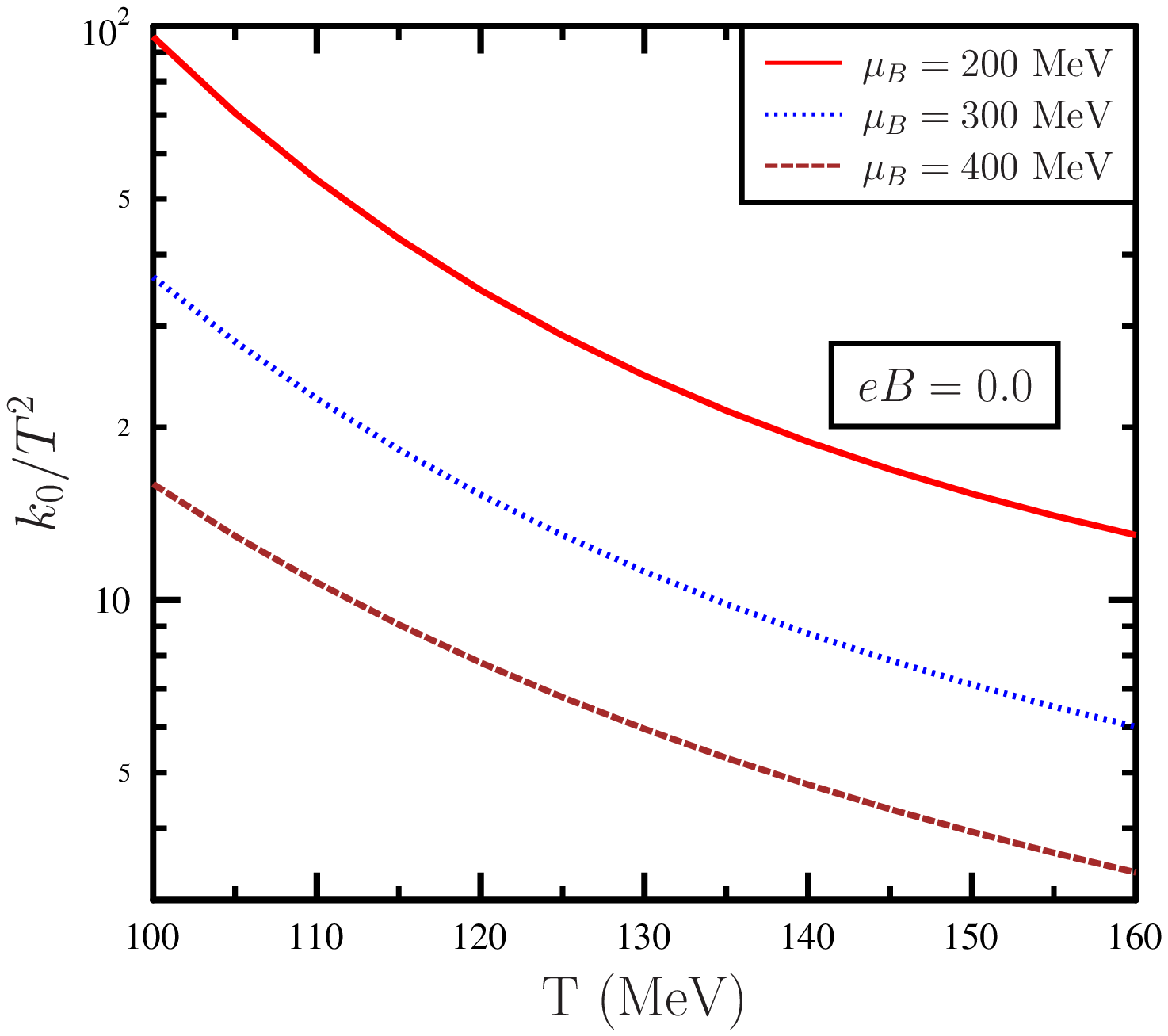}
\includegraphics[width=0.49\textwidth]{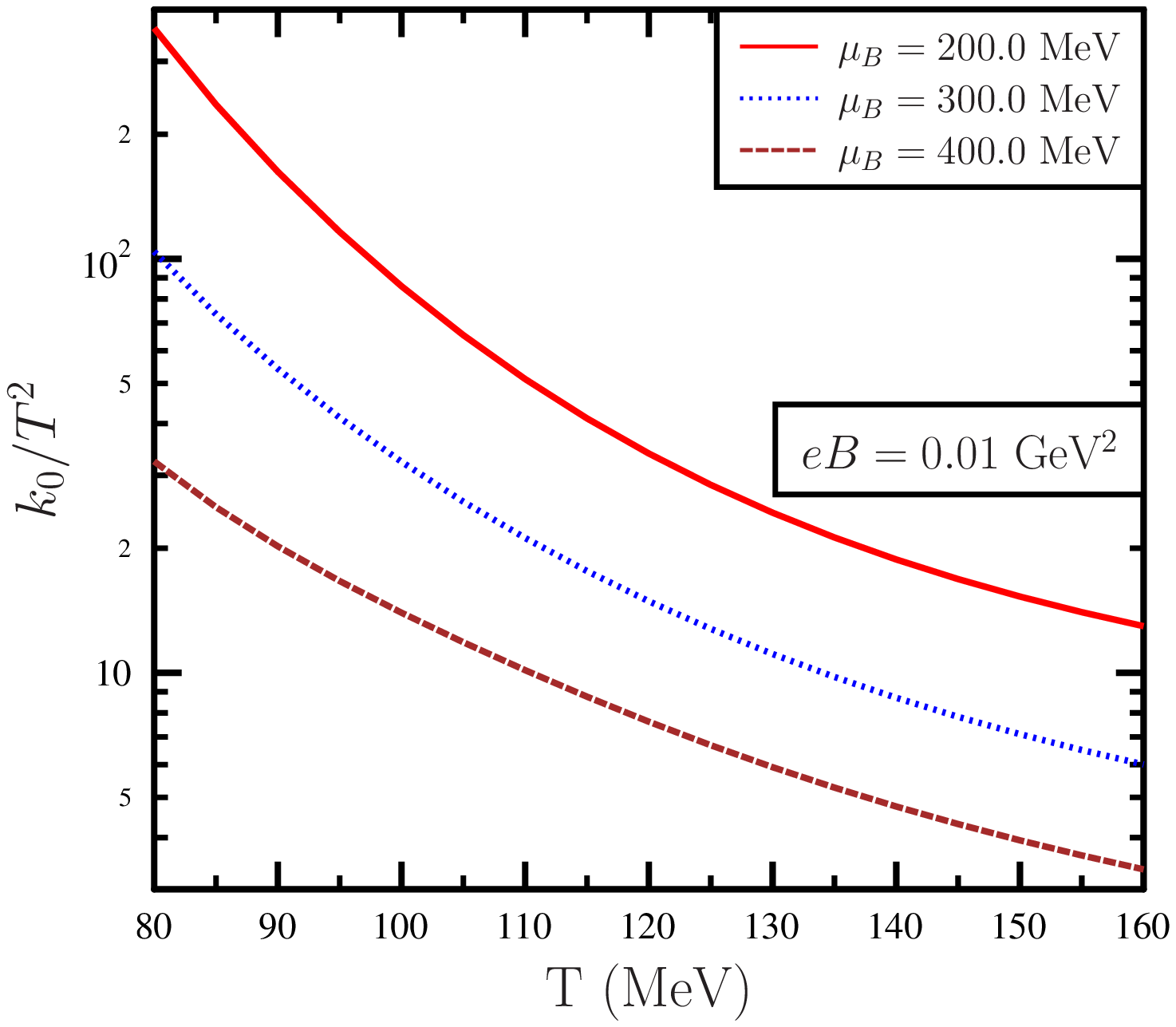}
\caption{Left plot: Variation of normalized thermal conductivity ($k_0/T^2$) with temperature ($T$) for different values of $\mu_B$  at zero magnetic field. Right plot:Variation of normalized thermal conductivity ($k_0/T^2$) with temperature ($T$) for different values of $\mu_B$ for a non vanishing magnetic field.  With increasing temperature and $\mu_B$, $k_0/T^2$ decreases. In the presence of magnetic field $k_0/T^2$ decrease.}
\label{fig2}
\end{center}
\end{figure}

We next discuss the variation of $k_0/T^2$ with $\mu_B$ in Fig.\eqref{fig2}. In the left panel we show the result for vanishing magnetic field and for non vanishing magnetic field on the right panel. With increasing $\mu_B$ and temperature, $k_0/T^2$ decreases. With increasing $\mu_B$ relaxation time of different hadrons and $\omega/n_B$ decreases. Relaxation time decreases with $\mu_B$, due to the fact that with increasing $\mu_B$ number density of the baryons increases. With increasing number density of the baryons scattering rate increases. On the other hand decreasing behaviour of $\omega/n_B$ with $\mu_B$ can be understood using Eq.\eqref{equ74}. From Eq.\eqref{equ74} it is clear that with increasing $\mu_B$, mesonic as well as baryonic contribution in $\omega/n_B$ of hadron resonance gas decreases due to the factor $\sinh(\mu_B/T)$ in the denominator. Mesonic contribution in energy density and pressure of hadron resonance gas is independent of $\mu_B$. On the other hand number density of baryons $(n_B)$ increases with $\mu_B$. Thus with increasing $\mu_B$
mesonic contribution in $\omega/n_B$ decreases. Further for baryons energy density, pressure and number density depend upon $\mu_B$. From Eq.\eqref{equ74} it is clear that energy density and pressure of baryons $\sim \cosh(\mu_B/T)$, but $n_B\sim \sinh(\mu_B/T)$. Hence with increasing $\mu_B$ baryonic contribution in $\omega/n_B$ of the hadron resonance gas decreases. Decreasing behaviour of $\omega/n_B$ with $\mu_B$ has been shown in Fig.\eqref{fig2new}.
For the range of temperature and $\mu_B$ considered in this investigation decrease of $\tau$ and $\omega/n_B$ is dominant with respect to increasing $f_0$ with $\mu_B$. Hence with $\mu_B$, $k_0/T^2$ of the hot and dense hadron gas decreases. It may be noted that as $\mu_B\rightarrow 0$ thermal conductivity diverges as $n_B^{-2}$. This divergence is inconsequential as the factor $k_0n_B^2$ enters the equation of motion. Since $k_0n_B^2$ remains finite as $\mu_B\rightarrow 0$, transport due to thermal conduction becomes irrelevant as $\vec{\nabla}(\mu_B/T)\rightarrow 0$ \cite{gavin1985}. For $\mu_B\rightarrow 0$ relevant transport processes are only momentum diffusion through viscous stresses. Such behaviour was also seen in Ref.\cite{paramitahm2016}
For non vanishing magnetic field $k_0/T^2$ decreases due to $(\omega_c\tau)^2$ factor in the denominator of expression for $k_0$.

\begin{figure}[!htp]
\begin{center}
\includegraphics[width=0.45\textwidth]{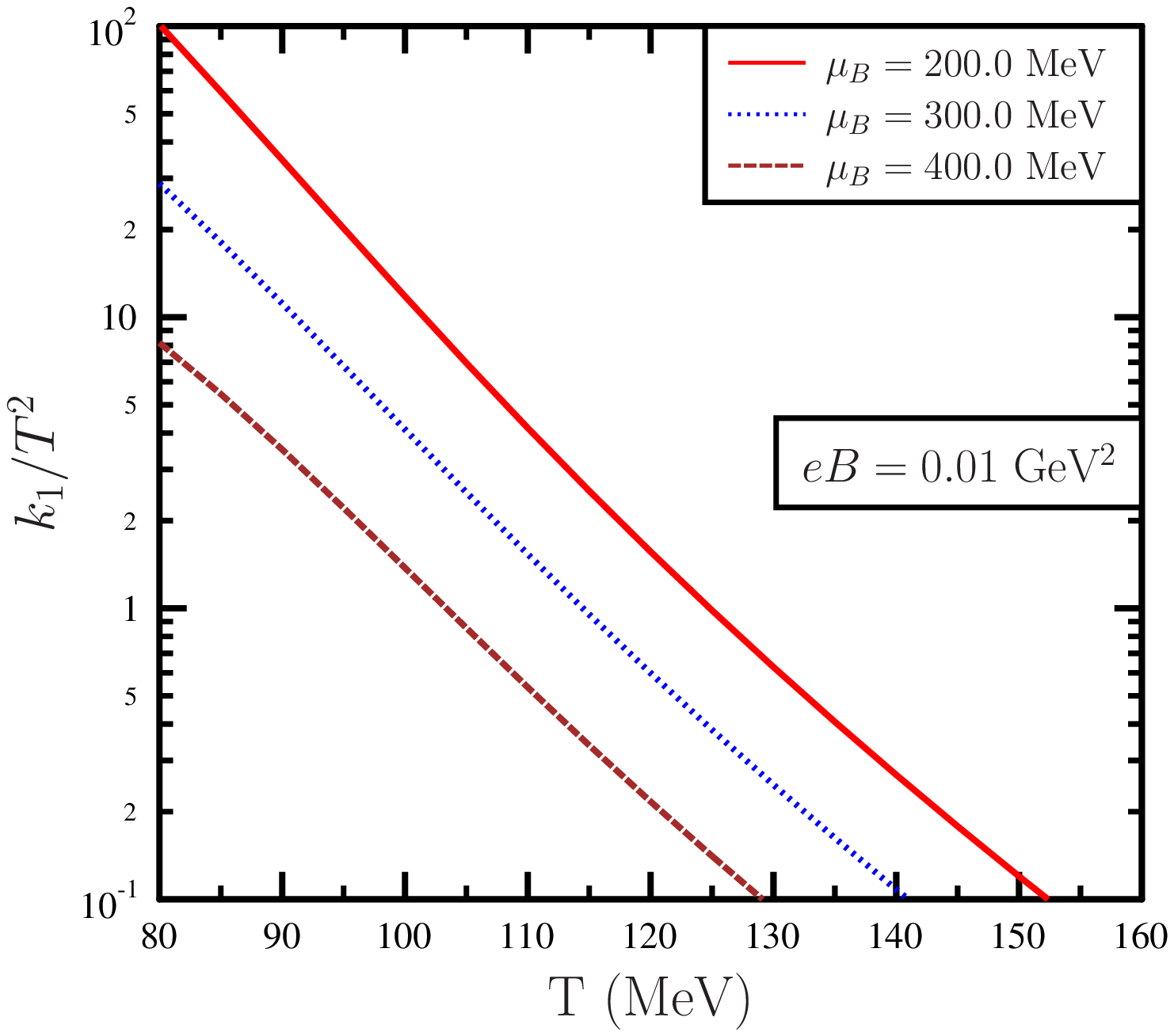}
\includegraphics[width=0.45\textwidth]{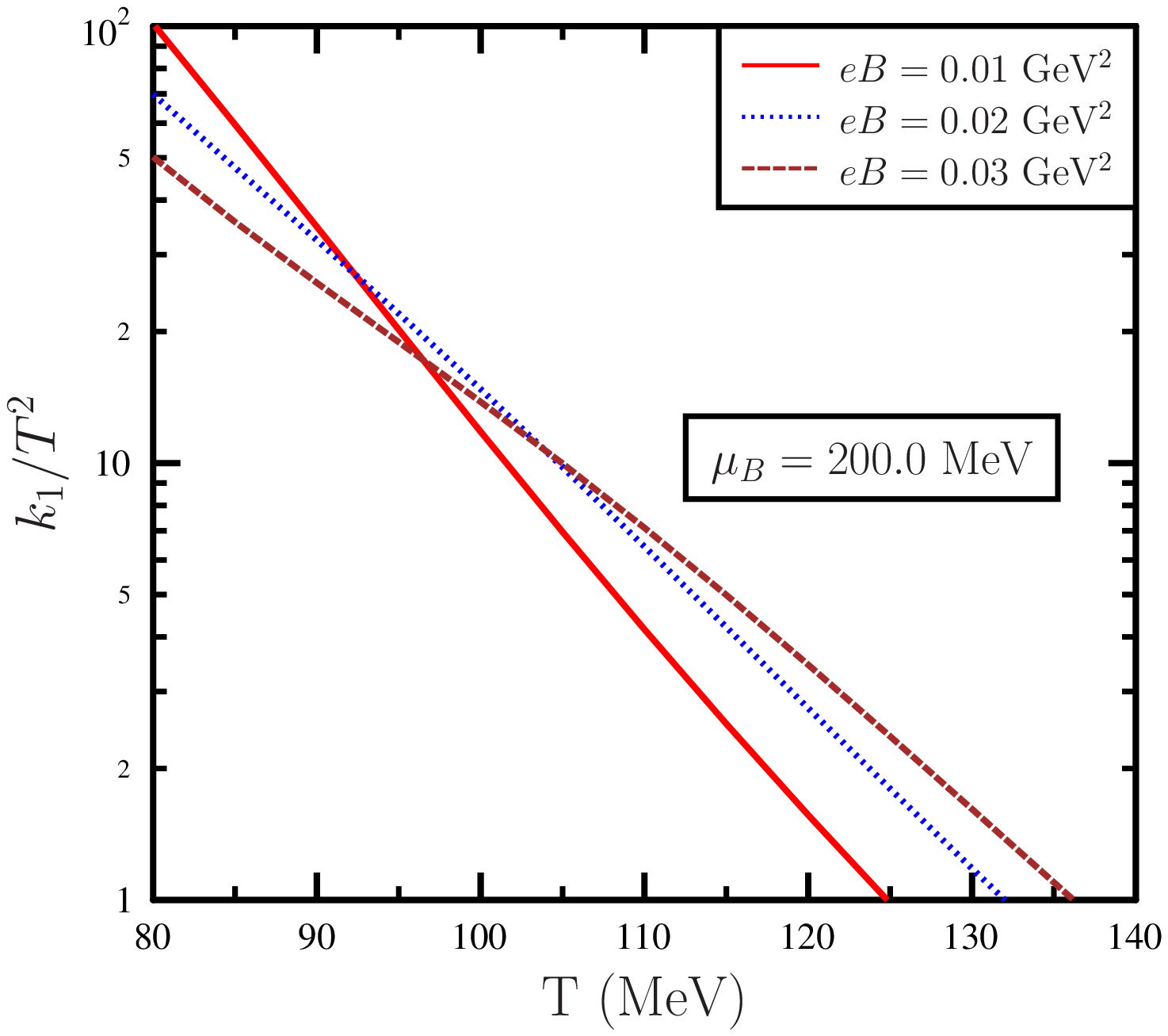}
\caption{Left Plot: Variation of Hall type thermal conductivity $k_1/T^2$ with temperature for non vanishing magnetic field and different values of $\mu_B$. Right Plot:
Variation of $k_1/T^2$ with temperature for different values of magnetic fields. With $\mu_B$ $k_1/T^2$ decreases. But for a fixed value of $\mu_B$ variation of $k_1/T^2$ is non monotonic with magnetic field. At low temperature $k_1/T^2$ decrease with magnetic field but at higher temperature $k_1/T^2$ increases with magnetic field.}
\label{fig4}
\end{center}
\end{figure}

In Fig.\eqref{fig4} we show the variation of Hall type thermal conductivity ($k_1/T^2$) with temperature. For vanishing magnetic field,  $k_1/T^2$ is zero as can be seen from Eq.\eqref{equ20}. Only at finite magnetic field and finite $\mu_B$, $k_1/T^2$ can have non vanishing values. In the left plot in Fig.\eqref{fig4} we show the variation of $k_1/T^2$ with temperature for non vanishing values of $\mu_B$ for a fixed value of magnetic field. It is clear from this plot that with $\mu_B$, $k_1/T^2$ decreases. This decrease is predominately due to decrease of $\omega/n_B$ factor with increasing $\mu_B$.
On the other hand for a fixed value of $\mu_B$, $k_1/T^2$ decreases with magnetic field at low temperature and increases with magnetic field at high temperature as can be seen in the right plot in Fig.\eqref{fig4}. This behavior of $k_1/T^2$ can be understood in the following way, at low temperature $\tau$ is large hence at low temperature $k_1/T^2\sim 1/\omega_c$. On the other hand at high temperature relaxation time is small and $k_1/T^2\sim \omega_c$. Thus variation of $k_1/T^2$ is different with magnetic field at low temperature and high temperature.

\begin{figure}[!htp]
\begin{center}
\includegraphics[width=0.45\textwidth]{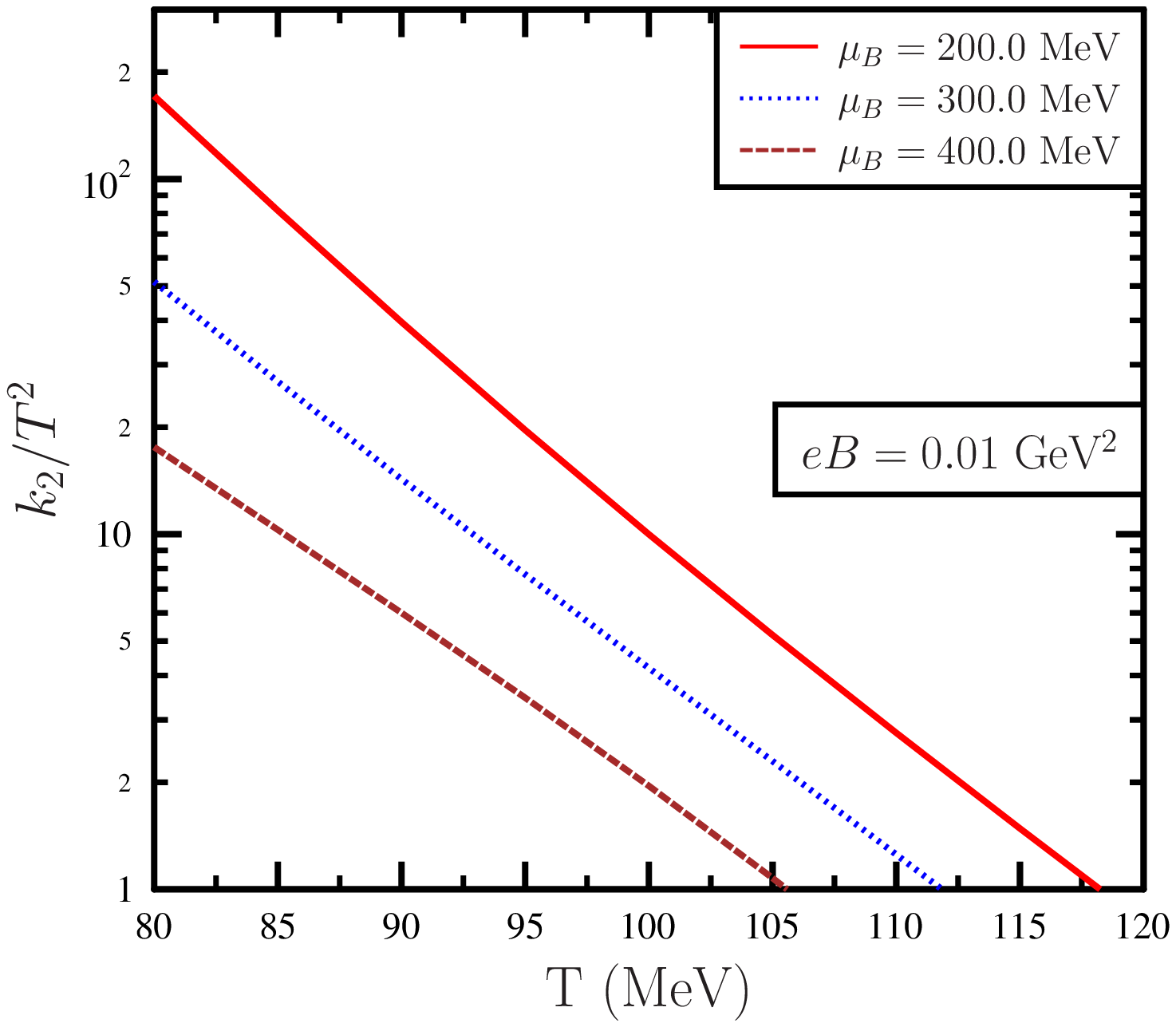}
\includegraphics[width=0.45\textwidth]{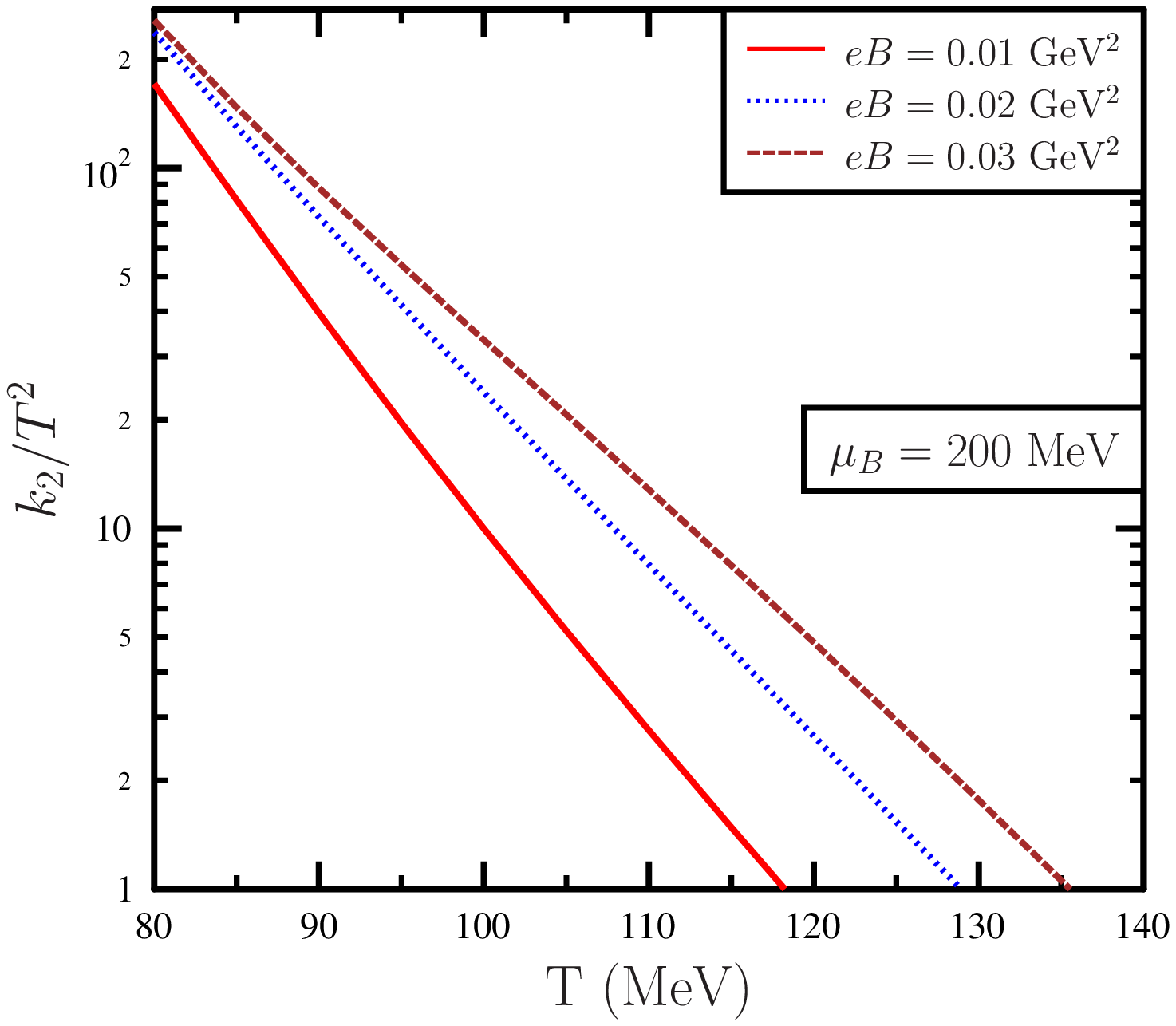}
\caption{Left plot: Variation of $k_2/T^2$ with temperature ($T$) for various values of baryon chemical potential at finite magnetic field. Right plot: Variation of $k_2/T^2$ with temperature ($T$) for various values of magnetic field. For a fixed value of magnetic field and $\mu_B$, $k_2/T^2$ decreases with temperature. With increasing magnetic field, generically $k_2/T^2$ increases within the range of $T$, $\mu_B$ and $B$ considered in this investigation. But with increasing $\mu_B$, $k_2/T^2$ decreases.}
\label{fig5}
\end{center}
\end{figure}

In Fig.\eqref{fig5} variation of the third component of the thermal conductivity tensor $k_2/T^2$ has been shown with temperature. It is clear for Eq.\eqref{equ21} that $k_2/T^2$ has non vanishing value only at finite magnetic field. In the right plot in Fig.\eqref{fig5} 
we show the variation of $k_2/T^2$ with temperature at non zero $\mu_B$ for various values of magnetic field. In the left plot in Fig.\eqref{fig5} we show the variation of $k_2/T^2$ with temperature with $\mu_B$ for non vanishing value of magnetic field. From the right plot in Fig.\eqref{fig5} we can see that with magnetic field $k_2/T^2$ increases. However for large magnetic field  and low temperature $k_2/T^2$ is not affected by magnetic field significantly. Naively this is because for low temperature the relaxation time is large hence $\frac{(\omega_c\tau)^2}{1+(\omega_c\tau)^2}\sim 1$ for high magnetic fields. Similarly at high temperature when the relaxation time is small, $k_2/T^2\sim \tau^2\omega_c^2$. Hence at high temperature with increasing magnetic field $k_2/T^2$ increases. In the left plot of Fig.\eqref{fig5} we show the variation of $k_2/T^2$ for non zero values of baryon chemical potential at finite magnetic field. In this plot we can see that with $\mu_B$, $k_2/T^2$ decreases. Variation of $k_2/T^2$ with $\mu_B$ is convoluted because in the expression of $k_2/T^2$  various terms are present which depend upon $\mu_B$, e.g. relaxation time, distribution function etc. With increasing $\mu_B$, both relaxation time as well as $\omega/n_B$ decreases and $f_0$
increases. But the increase of $f_0$ with $\mu_B$ is not large enough to compensate the decreasing behavior of $\tau$ and $\omega/n_B$. Hence with increasing $\mu_B$, $k_2/T^2$ decrease.

\subsection{Results for electrical conductivity in a magnetic field}

\begin{figure}[!htp]
\begin{center}
\includegraphics[width=0.45\textwidth]{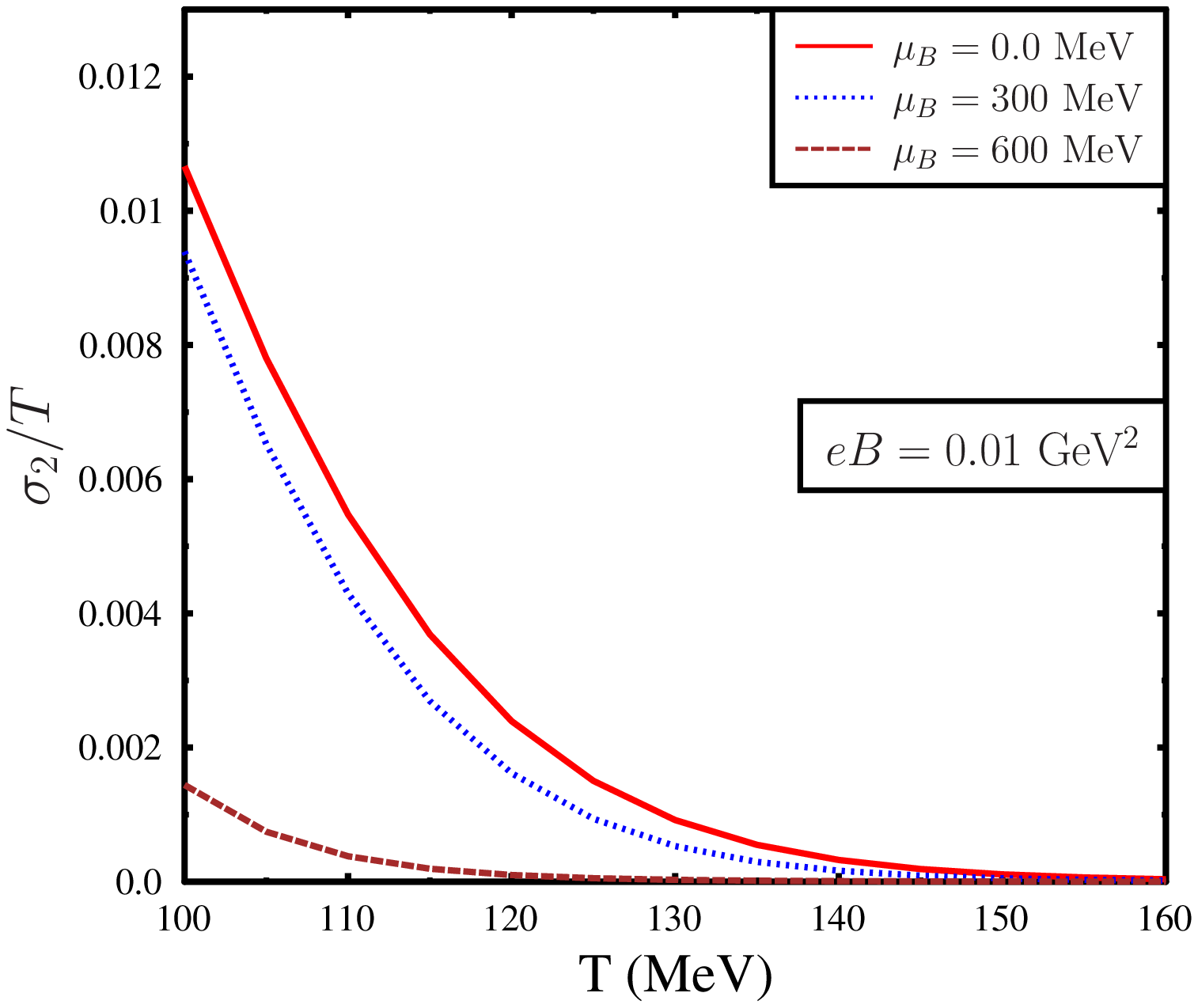}
\includegraphics[width=0.45\textwidth]{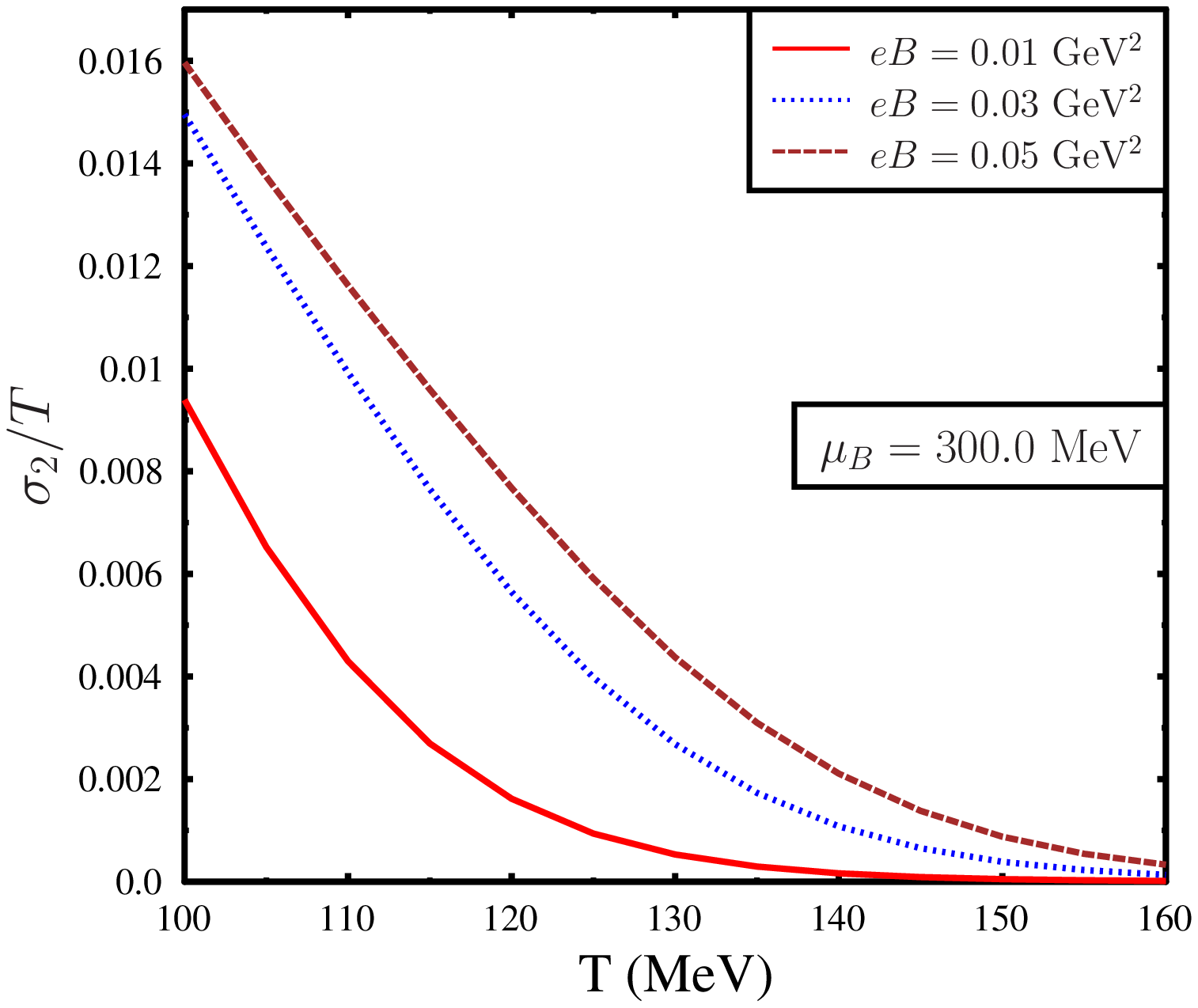}
\caption{Left plot: Variation of $\sigma_2/T$ with temperature for a non vanishing magnetic field and for different values of $\mu_B$. With increasing $\mu_B$ $\sigma_2/T$ decreases. Right plot: Variation of $\sigma_2/T$ with temperature for non vanishing values of magnetic field at finite $\mu_B$. With magnetic field $\sigma_2/T$ increases.}
\label{fig6}
\end{center}
\end{figure}

In this subsection we discuss the variation of electrical conductivity ($\sigma_2/T$) with temperature, magnetic field and baryon chemical potential. In our earlier work we had demonstrated in details variation of $\sigma_0$ and $\sigma_1$ with temperature, magnetic field and $\mu_B$\cite{hallhrg}. Therefore we do not repeat the discussion on the results for $\sigma_0$ and $\sigma_1$ here. Here we only show the variation of $\sigma_2$ with temperature, baryon chemical potential and magnetic field. In the left plot of Fig.\eqref{fig6} we show the variation of $\sigma_2/T$ with temperature for a non vanishing value of magnetic field but with different values of $\mu_B$. For a fixed value of magnetic field and $\mu_B$  $, \sigma_2/T$ decreases with temperature. Among various hadrons mesonic contribution to $\sigma_2/T$ is large with respect to the baryonic contribution. With increasing temperature mesonic contribution decreases due to decrease in the relaxation time of mesons. With increasing $\mu_B$ mesonic contribution to $\sigma_2/T$ decreases and the baryonic contribution increases. However the decrease in mesonic contribution with increasing $\mu_B$ is not compensated with increasing baryonic contribution for the range of temperature and baryon chemical potential considered here. Hence with increasing $\mu_B$, $\sigma_2/T$ decreases.  Right plot in Fig.\eqref{fig6} we show the variation of $\sigma_{2}/T$ with magnetic field. For the range of $T$, $\mu_B$ and $B$ we considered in this investigation $\sigma_2/T$ increases with magnetic field.    

\subsection{Results for shear viscosity in a magnetic field}

\begin{figure}[!htp]
\begin{center}
\includegraphics[width=0.45\textwidth]{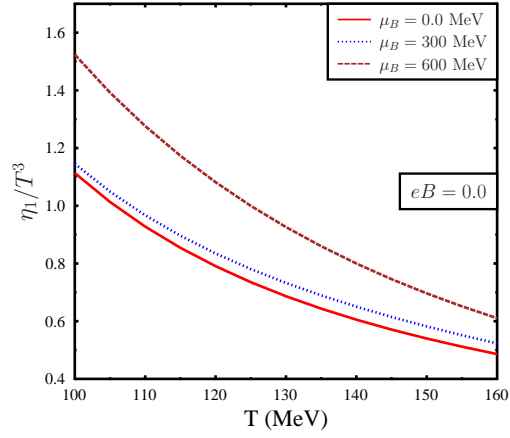}
\caption{Variation of $\eta_1/T^3$ with temperature for vanishing magnetic field but with different values of $\mu_B$. At zero magnetic field $\eta_1=\eta_2$. From this figure it is clear that $\eta_1/T^3$ decreases with temperature and increases with $\mu_B$.}
\label{fig7}
\end{center}
\end{figure}

In Fig.\eqref{fig7} we show the variation of $\eta_1/T^3$ with temperature for non vanishing values of $\mu_B$ for zero magnetic field. It is important to note that for zero magnetic field $\eta_1=\eta_2$ and $\eta_3=\eta_4=0$.
With temperature $\eta_1/T^3$ decrease and it increases with $\mu_B$. This behavior of $\eta_1/T^3$ with $\mu_B$ is a combined effect of the variation of relaxation time and distribution function with $\mu_B$. Relaxation time decrease with $\mu_B$, however with increasing baryon chemical potential $f_0$ increases. Among various hadrons mesonic contribution to $\eta_1/T^3$ 
is large with respect to the baryonic contribution at zero $\mu_B$. With increasing temperature relaxation time of the hadrons decreases which gives rise to the decreasing behaviour of $\eta_1/T^3$ with temperature. On the other hand with $\mu_B$ mesonic contribution decreases due to decrease in relaxation time with $\mu_B$, however with $\mu_B$ baryonic contribution increases due to the $\mu_B$ factor in the distribution function. This increasing contributions of baryons at finite $\mu_B$ compensate decreasing contributions of the mesons. With increasing $\mu_B$ 
baryonic contribution becomes significant over mesonic contribution. Hence with $\mu_B$, $\eta_1/T^3$ as well as $\eta_2/T^3$ increases.

\begin{figure}[!htp]
\begin{center}
\includegraphics[width=0.45\textwidth]{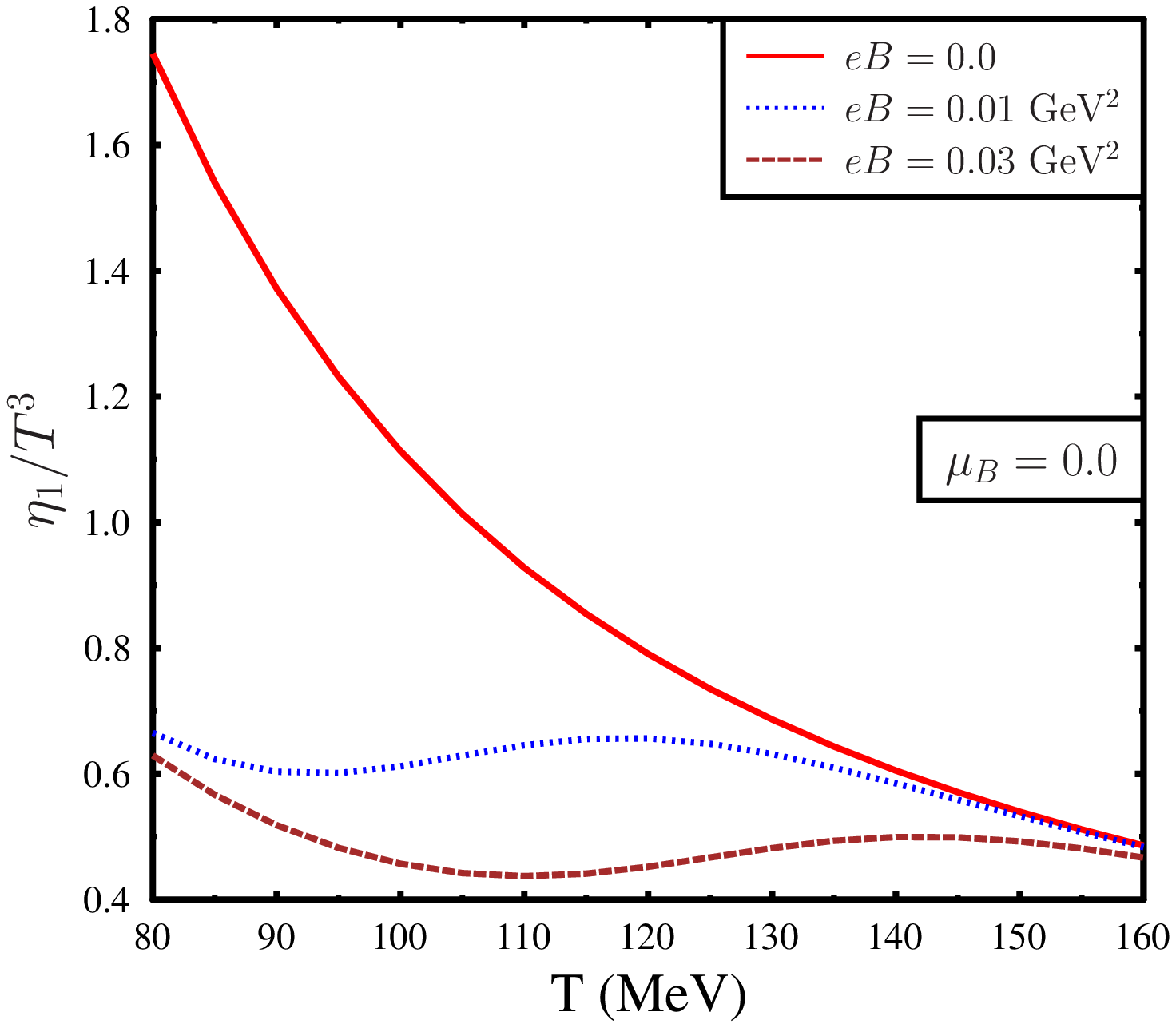}
\includegraphics[width=0.45\textwidth]{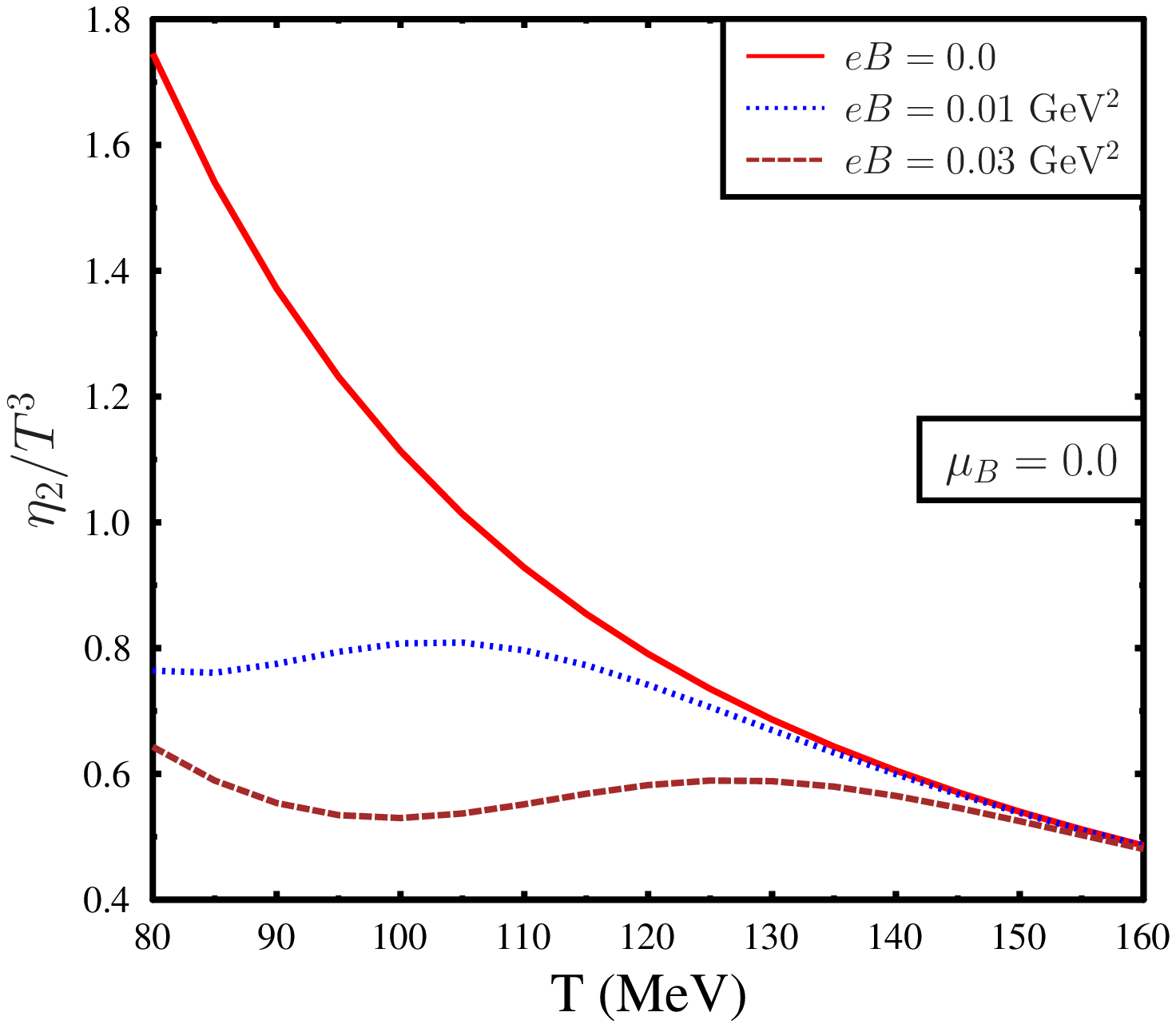}
\caption{Left plot: Variation of $\eta_1/T^3$ with temperature for vanishing $\mu_B$ but with different values of magnetic field. Right plot: Variation of $\eta_2/T^3$  with temperature for vanishing $\mu_B$ but with different values of magnetic field. For vanishing magnetic field $\eta_1=\eta_2$ which can be seen in this figure. Also for non vanishing magnetic field $\eta_1\neq\eta_2$ as can be seen in these plots. With increasing magnetic field both $\eta_1/T^3$ and $\eta_2/T^3$ decreases. This decrease   is very prominent at low temperature. At high temperature effect of magnetic field is not significant.}
\label{fig8}
\end{center}
\end{figure}

In Fig.\eqref{fig8} we show the variation of $\eta_1/T^3$ and $\eta_2/T^3$ with temperature for non vanishing values of magnetic field at zero baryon chemical potential. From the 
expression of $\eta_1$ and $\eta_2$ it is clear that in the presence of magnetic field their behavior is similar. $\eta_1$ and $\eta_2$ only differs in numerical values.
From the left plot in Fig.\eqref{fig8} we can see that with magnetic field $\eta_1/T^3$ decreases. This is due to $(\omega_c\tau)^2$ factor in the denominator of Eq.\eqref{equ56}.
However for a fixed value of magnetic field variation of $\eta_1/T$ with temperature is rather non monotonic in nature. Variation of $\eta_1/T$ with temperature for a non vanishing value of magnetic field shows a peak structure. It is also important to note that at high temperature magnetic field does not affect $\eta_1/T^3$ significantly. This is because at high temperature relaxation time is small hence the factor $\omega_c\tau$ in the denominator of Eq.\eqref{equ56} is rather small. Hence at high temperature suppression effect due to magnetic field is not significant. In the right plot of Fig.\eqref{fig8} we show the variation of $\eta_2/T$ with temperature for non vanishing values of magnetic field at zero $\mu_B$. Behaviour of $\eta_2/T$ can also be understood in the same manner as we have discussed for  $\eta_1/T$.

\begin{figure}[!htp]
\begin{center}
\includegraphics[width=0.45\textwidth]{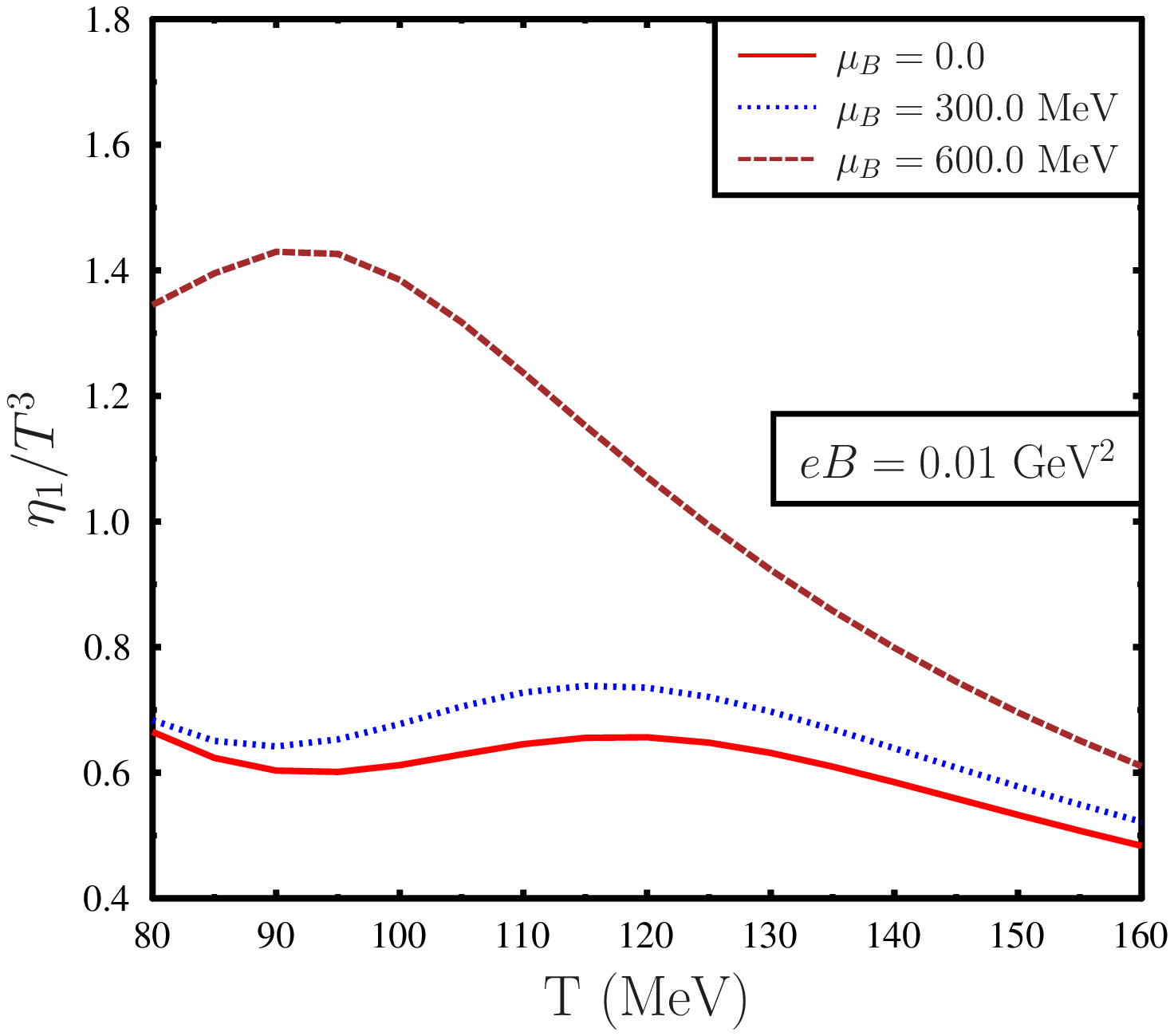}
\includegraphics[width=0.45\textwidth]{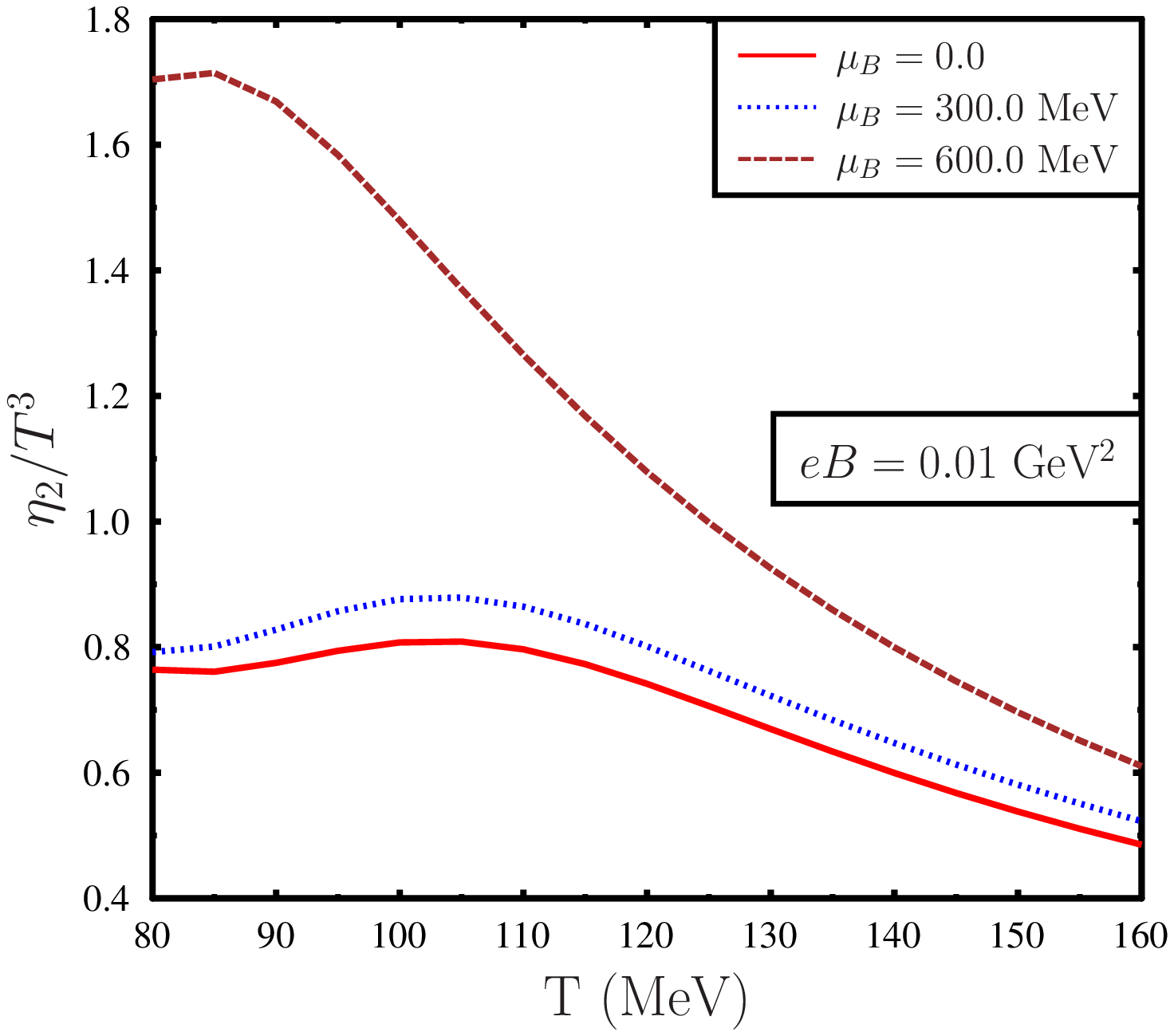}
\caption{Left plot: Variation of $\eta_1/T^3$ with temperature for different values of $\mu_B$ in the presence of magnetic field. 
Right plot: Variation of $\eta_2/T^3$ with temperature for different values of $\mu_B$ in the presence of magnetic field. Behaviour of $\eta_1/T^3$ and $\eta_2/T^3$ are similar apart from their numerical values. $\eta_2$ is larger than $\eta_1$ as can be seen from their analytical expressions. For higher $\mu_B$ value of $\eta_1/T^3$ and $\eta_2/T^3$ is higher. Variation of both $\eta_1/T^3$ and $\eta_2/T^3$ with temperature shows non monotonic behavior with a peak structure.}
\label{fig9}
\end{center}
\end{figure}

Next we show the variation of $\eta_1/T^3$ and $\eta_2/T^3$ with temperature for a non vanishing magnetic field and various values of $\mu_B$ in Fig.\eqref{fig9}. In the presence of magnetic field $\eta_1$ is smaller than $\eta_2$ as can be seen from Eq.\eqref{equ56} and Eq.\eqref{equ57}. This apart variation of $\eta_1/T^3$ and $\eta_2/T^3$ are similar with temperature, magnetic field and $\mu_B$. From Fig.\eqref{fig9} it is clear that variation of $\eta_1/T^3$ and $\eta_2/T^3$ with $\mu_B$ is similar to Fig.\eqref{fig7} i.e. with $\mu_B$,
$\eta_1/T^3$ and $\eta_2/T^3$ increases. Mesonic contribution to $\eta_1/T^3$ and $\eta_2/T^3$ is significantly larger than the baryonic contribution at vanishing $\mu_B$. With increasing $\mu_B$ mesonic contribution decreases due to decrease in the relaxation time of mesons. On the other hand with increasing $\mu_B$ baryonic contribution increases due to $\mu_B$ factor in the distribution function. Increasing baryonic contribution compensate decreasing mesonic contribution to $\eta_1/T^3$ and $\eta_2/T^3$. Hence with increasing $\mu_B$ both $\eta_1/T^3$ and $\eta_2/T^3$ increases. However in the presence of magnetic field values of $\eta_1$ and $\eta_2$ are smaller with respect to the same in the absence of magnetic field. For a non vanishing value of $\mu_B$ and magnetic field variation of $\eta_1/T^3$ and $\eta_2/T^3$ with temperature is non monotonic and is similar to Fig.\eqref{fig8}.

\begin{figure}[!htp]
\begin{center}
\includegraphics[width=0.45\textwidth]{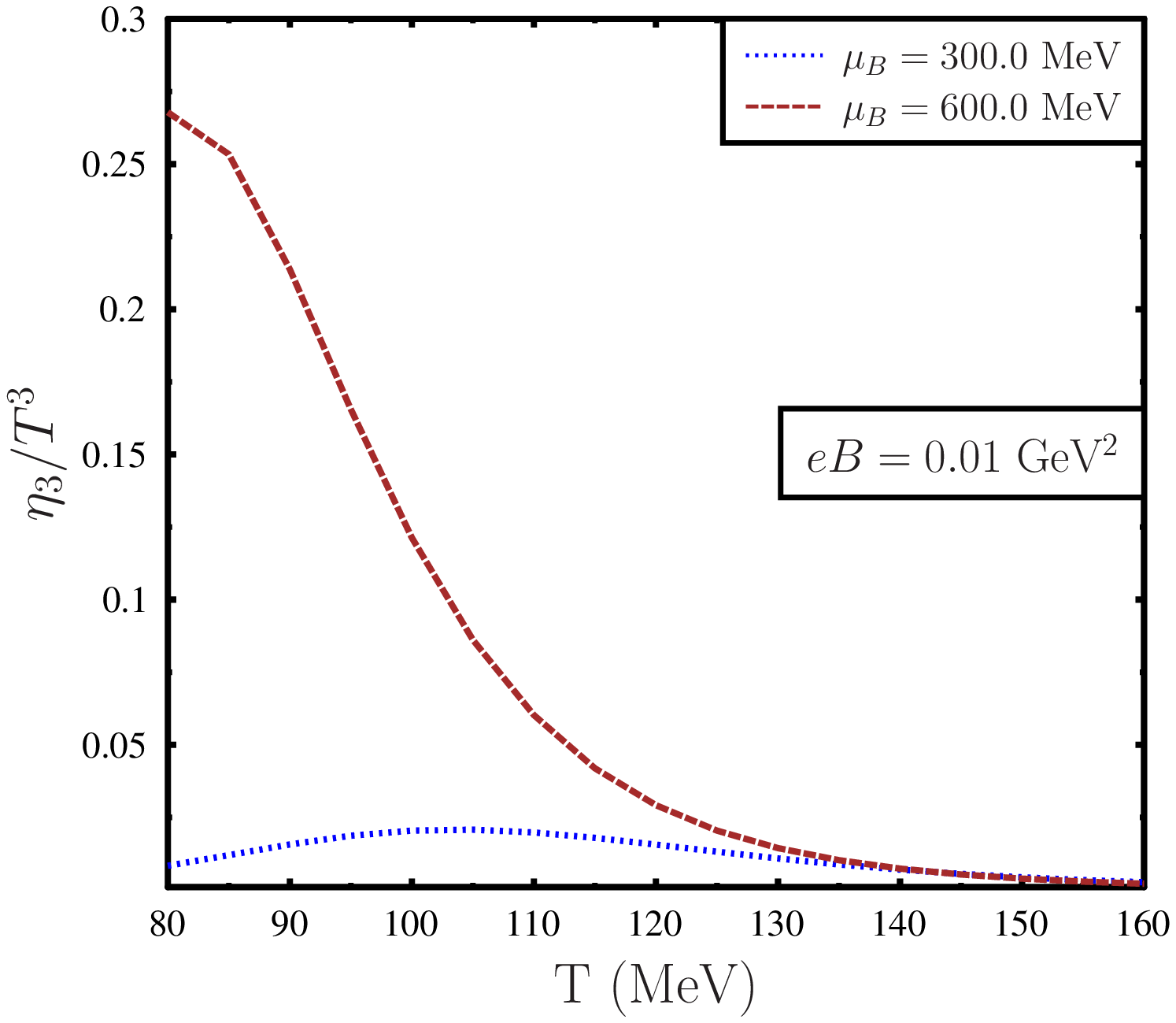}
\includegraphics[width=0.45\textwidth]{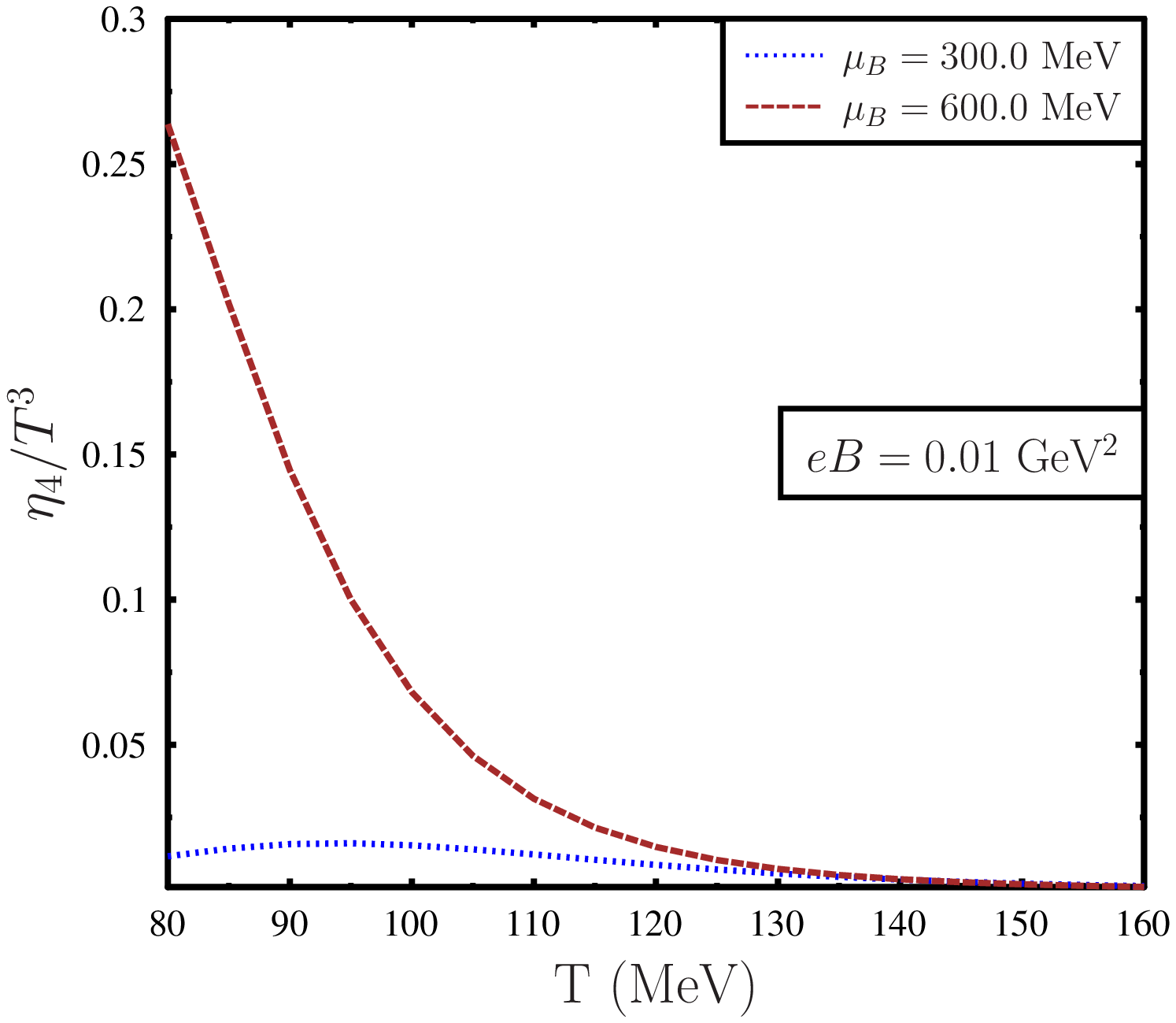}
\caption{Left plot: Variation of $\eta_3/T^3$ with temperature for a non vanishing magnetic field and different values of $\mu_B$. Right plot:Variation of $\eta_4/T^3$ with temperature for a non vanishing magnetic field and different values of $\mu_B$. Behavior of $\eta_3/T^3$ and $\eta_4/T^3$ are similar with $\mu_B$ and temperature apart from their numerical values. With increasing $\mu_B$ hall type shear viscosities $\eta_3/T^3$ and $\eta_4/T^3$ increases.}
\label{fig10}
\end{center}
\end{figure}

In Fig.\eqref{fig10} we show the variation of $\eta_3/T^3$ and $\eta_4/T^3$ with temperature for non vanishing values of $\mu_B$ for a finite magnetic field. Note that $\eta_3$ and $\eta_4$ are Hall type shear viscosities in magnetic field. Hence $\eta_3$ and $\eta_4$ are zero for zero magnetic field as well as for zero baryon chemical potential due to equal and opposite contributions of particles and antiparticles. Only at non vanishing magnetic field and finite $\mu_B$, $\eta_3$ and $\eta_4$ has non vanishing values. From Fig.\eqref{fig10} we see that with $\mu_B$ both $\eta_3/T^3$ and $\eta_4/T^3$ increases for non vanishing value of magnetic field. This behavior of $\eta_3$ and $\eta_4$ can be understood naively in the following way. Due to Hall type nature of $\eta_3$ and $\eta_4$ only baryons contribute to  $\eta_3$ and $\eta_4$ at finite $\mu_B$. With increasing $\mu_B$ number density of baryons increases, this gives rise to increasing behaviour of $\eta_3/T^3$ and $\eta_4/T^3$.

\begin{figure}[!htp]
\begin{center}
\includegraphics[width=0.45\textwidth]{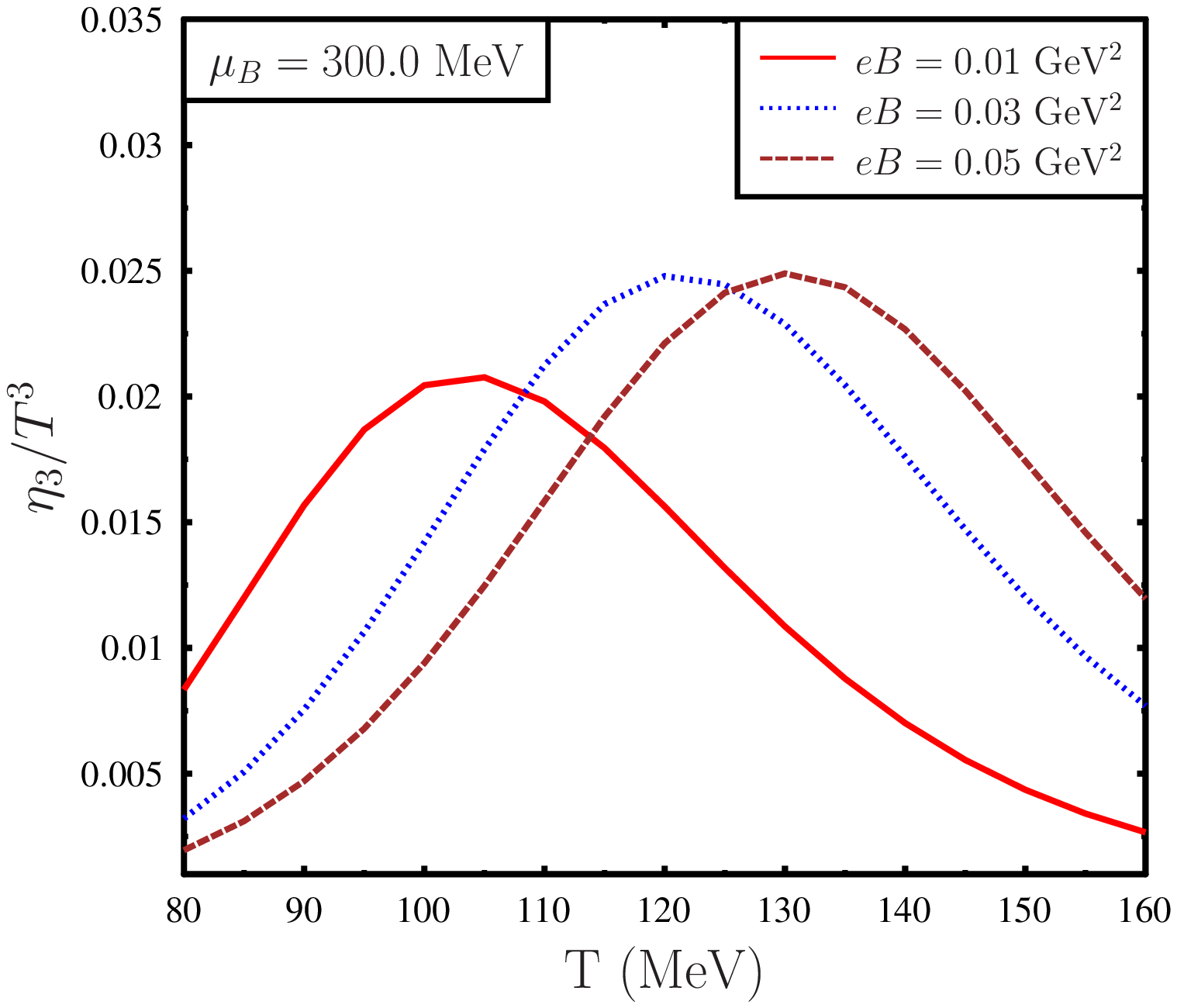}
\includegraphics[width=0.45\textwidth]{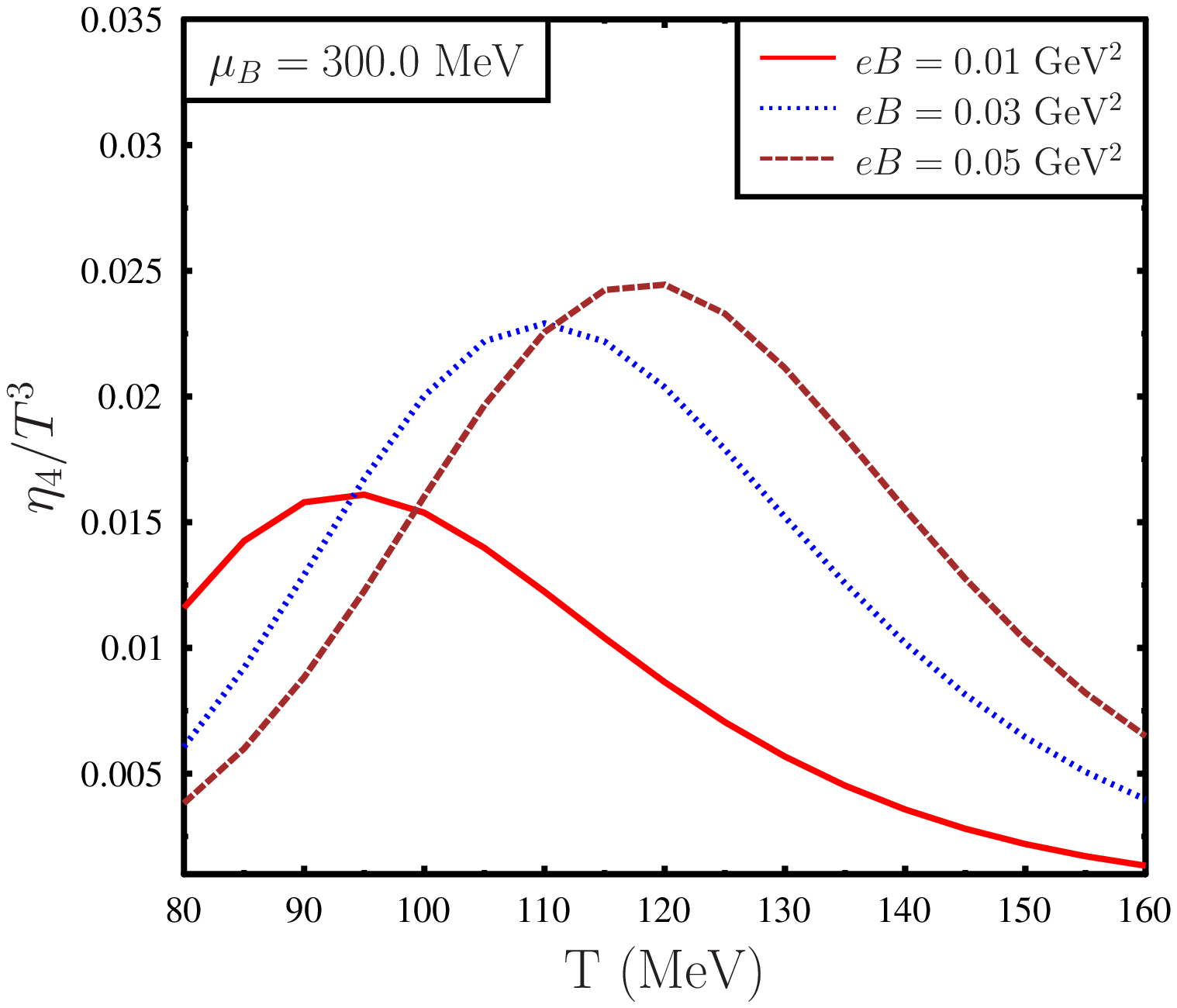}
\caption{Left plot: Variation of $\eta_3/T^3$ with temperature for different values of non vanishing magnetic field and finite $\mu_B$.
Right plot: Variation of $\eta_4/T^3$ with temperature for different values of non vanishing magnetic field and finite $\mu_B$. Variation of $\eta_3/T^3$ and $\eta_4/T^3$ with temperature and magnetic field are similar apart from their numerical values. For low temperature $\eta_3/T^3$ and $\eta_4/T^3$ decreases with magnetic field. On the other hand for high temperature both $\eta_3/T^3$ and $\eta_4/T^3$ increases with magnetic field.
Variation of $\eta_3/T^3$ and $\eta_4/T^3$ with temperature is non monotonic with a peak.}
\label{fig11}
\end{center}
\end{figure}

In Fig.\eqref{fig11} we show the variation of $\eta_3/T^3$ and $\eta_4/T^3$ with temperature for non vanishing values of magnetic field at finite $\mu_B$. From this figure we can see that with increasing magnetic field both $\eta_3/T^3$ and $\eta_4/T^3$ increases at large temperature. But at low temperature both $\eta_3/T^3$ and $\eta_4/T^3$ decreases with magnetic field. This behavior of $\eta_3/T^3$ and $\eta_4/T^3$ is similar to the other Hall type conductivities as discussed earlier. 
At low temperature due to large value of relaxation time both $\eta_3/T^3$ and $\eta_4/T^3\sim 1/\omega_c$. On the other hand at high temperature due to large relaxation time $\eta_3/T^3\sim \omega_c$. This different behavior of $\eta_3/T^3$ as well as $\eta_4/T^3$ with magnetic field at high and low temperature give rise to non monotonic variation of Hall type shear viscosities with magnetic field.

\section{conclusion}

Off central heavy ion collisions can produce a strong magnetic field. The life time of such a field during the evolution of
QGP to hadron gas depends critically on transport coefficients like electrical conductivity. Similarly the other dissipative
coefficients in presence of magnetic field are also important and essential ingredients for the magnetohydrodynamic evolution
of the strongly interacting medium produced subsequent to the collision. We have here attempted to evaluate some of these
coefficients for magnetized hot and dense hadronic matter. The explicit calculations are performed within the hadron resonance gas model.

We have used the Boltzmann transport equation in the relaxation time approximation to estimate the transport coefficients.
We have incorporated the effect of magnetic field through the cyclotron frequency of individual hadrons. Due to the vector
nature of the field, the transport coefficients are no longer isotropic. 
It is observed that the anisotropic transport coefficients are always smaller than their isotropic counterpart at vanishing magnetic
field. For strong fields the effects arising from collision become smaller compared to the effects arising from the cyclotron frequency.

For shear viscosity due to the presence of magnetic field, the transverse viscosity coefficient will be smaller compared to the
longitudinal viscosity coefficient and will affect transverse flow.
The structure of the viscous stress tensor in a magnetic field is  model independent.  However the precise value of the transverse
shear viscosity depend on the model considered. The viscous properties of the fluid extracted from flow data can lead to a more ideal 
fluid behaviour in the presence of magnetic field as compared to the case in the absence of magnetic field \cite{tuchin3}.

In the context of electrical conductivity, it was seen that Hall conductivity $\sigma_1$ for hadron gas generically increases with
magnetic field \cite{hallhrg}. It is also observed here that the non Hall type conductivity $\sigma_2$ increases with magnetic field
while the component $\sigma_0$ decreases with magnetic field. It is to be noted that $\sigma=\sigma_0+\sigma_2$ is the electrical
conductivity in the absence of magnetic field. 

Similar behavior is also observed for the anisotropic thermal conductivities ($k_0,k_1$ and $k_2$). The ``Hall type'' thermal conductivity $k_1$ generically increases with magnetic field. The non Hall type conductivity $k_0$ decreases with magnetic field while $k_2$ increases with magnetic field keeping $k=k_0+k_2$, being independent of magnetic field with a value as one would obtain in the absence of magnetic field. 
In the present work, we have included the effect of magnetic field through cyclotron frequency of individual hadrons and have not taken the quantum effects arising from Landau quantization. Further the relaxation time has been included with a hard sphere scattering where effect of magnetic field is not included. Some of these calculations are in progress and will be reported elsewhere.

\section*{Acknowledgement}
RKM would like to thank PRL for support and local hospitality for her visit, during which this problem was initiated.
 RKM would like to thank Basanta K. Nandi and Sadhana Dash for constant support and encouragement.


\begin{thebibliography}{99}
\bibitem{HeinzSnellings2013}
U. W. Heinz and R. Snellings, Annu. Rev. Nucl. Part. Sci. {\bf 63},
123-151 (2013).
\bibitem{RomatschkeRomatschke}
P. Romatschke and U. Romatschke, Phys.Rev.Lett. {\bf 99},172301 (2007).
\bibitem{KSS}
P.K. Kovtun, D.T.Son and A.O.Starinets, Phys.Rev.Lett. {\bf 94},111601 (2005).
\bibitem{gavin1985}
S. Gavin, Nucl. Phys. {\bf A 435}, 826 (1985).
\bibitem{kajantie1985}
A. Hosoya, K. Kajantie, Nucl. Phys. {\bf B250}, 666 (1985).
\bibitem{DobadoTorres2012}
A.Dobado and J. M. Torres-Rincon, Phys. Rev.{\bf D86}, 074021 (2012).
\bibitem{sasakiRedlich2009}
C. Sasaki and K.Redlich, Phys. Rev.{\bf C79}, 055207 (2009).
\bibitem{sasakiRedlich2010}
C. Sasaki and K.Redlich, Nucl. Phys.{\bf A832}, 62 (2010).
\bibitem{KarschKharzeevTuchin2008}
F. Karsch, D. Kharzeev, and K. Tuchin, Phys. Lett. {\bf B663}, 217 (2008).
\bibitem{FinazzoRougemont2015}
S. I. Finazzo, R. Rougemont, H. Marrochio, J. Noronha, JHEP 1502, 051 (2015).
\bibitem{WiranataPrakash2009}
A. Wiranata and M. Prakash, Nucl. Phys.{\bf A830}, 219C-222C (2009) 
\bibitem{JeonYaffe1996}
S. Jeon and L. Yaffe, Phys.Rev.{\bf D53}, 5799-5809 (1996).
\bibitem{mclerran2008}
D. E. Kharzeev, L. D. McLerran, H. J. Warringa, Nucl. Phys. {\bf A 803}, 227 (2008).
\bibitem{skokov}
V. Skokov, A. Yu. Illarionov, V. Toneev, Int. J. Mod. Phys. {\bf A 24}, 5925 (2009).
\bibitem{kharzeevbook}
``Strongly interacting Matter in Magnetic field'', edited by D. Kharzeev, K. Landsteiner, A. Schmitt 
and H. Yee, Lecture Notes in Physics vol 871, Springer-Verlag Berlin Heidelberg 2013. 

\bibitem{MHD1}
G. Inghirami et al., Eur. Phys. J. C 76,659 (2016).
\bibitem{MHDajit}
A. Das, S.S. Dave, P.S. Saumia, A.M. Srivastava, Phys.Rev.{\bf C96}, 034902 (2017).
\bibitem{TuchinMHD}
K.Tuchin, Phys.Rev.{\bf C83}, 017901 (2011); Phys. Rev.{\bf C82}, 034904 (2010).
\bibitem{MoritzGreif} 
M. Greif, C. Greiner, and G. S. Denicol, Phys.Rev. {\bf D93} (2016) no.9, 096012.
\bibitem{electricalcond1}
 M. Greif, I. Bouras, C. Greiner, and Z. Xu, Phys. Rev. {\bf D90}, 094014 (2014).
\bibitem{electricalcond2} 
 A. Puglisi, S. Plumari, and V. Greco, arXiv:1407.2559.
 \bibitem{electricalcond3}
 A. Puglisi, S. Plumari, and V. Greco, Phys. Rev. {\bf D90}, 114009 (2014).
 \bibitem{electricalcond4}
  W. Cassing, O. Linnyk, T. Steinert, and V. Ozvenchuk, Physical Review Letters 110, 182301 (2013).
  \bibitem{electricalcond5}
  T. Steinert and W. Cassing, Physical Review C 89, 035203 (2014).
  \bibitem{electricalcond6}
   G. Aarts, C. Allton, A. Amato, P. Giudice, S. Hands, and J.-I. Skullerud, JHEP 02, 186 (2015).
  \bibitem{electricalcond7}
   G. Aarts, C. Allton, J. Foley, S. Hands, and S. Kim, Physical Review Letters 99, 022002 (2007).
   \bibitem{electricalcond8}
    A. Amato, G. Aarts, C. Allton, P. Giudice, S. Hands,  and J.-I. Skullerud, Physical Review Letters 111, 172001 (2013).
 \bibitem{electricalcond9}
  S. Gupta, Physics Letters B 597, 57 (2004).
  \bibitem{electricalcond10}
   Y. Burnier and M. Laine, The European Physical Journal C 72, 1902 (2012).
   \bibitem{electricalcond11}
    H.-T. Ding, A. Francis, O. Kaczmarek, F. Karsch, E. Laermann, and W. Soeldner, Physical Review D 83, 034504 (2011).
\bibitem{electricalcond12}
 O. Kaczmarek and M. M̈uller, PoS LATTICE2013 , 175 (2014)
\bibitem{electricalcond13}
 S.-X. Qin, Phys. Lett. B742, 358 (2015).
\bibitem{electricalcond14}
 R. Marty, E. Bratkovskaya, W. Cassing, J. Aichelin, and H. Berrehrah, Phys. Rev. C88, 045204 (2013)
 \bibitem{electricalcond15}
  D. Fern ́andez-Fraile and A. Gomez Nicola, Physical Review D 73, 045025 (2006). 
  \bibitem{danicol2014}
G.S. Denicol, H. Niemi, I. Bouras E. Molnar , Z. Xu , D.H. Rischke, C. Greiner ,Phys. Rev. {\bf D 89}, 074005 (2014).
\bibitem{Kapusta2012}
J.I. Kapusta and J.M. Torres-Rincon,Phys. Rev. {\bf C86}, 054911 (2012).

 \bibitem{danicol2018}
 M. Greif, J. A. Fotakis, G. S. Denicol, C. Greiner, Phys. Rev. Lett. {\bf 120}, 242301 (2018). 
 \bibitem{PrakashVenu}
 M. Prakash, M. Prakash, R. Venugopalan and G. Welke, Phys.Rept.{\bf 227}, 321-366 (1993).
\bibitem{WiranataPrakash2012}
A. Wiranata and Madappa Prakash, Phys. Rev.{\bf C85}, 054908 (2012).
\bibitem{KapustaChakraborty2011}
 P. Chakraborty and J.I. Kapusta Phys. Rev.{\bf C83}, 014906 (2011).
\bibitem{Toneev2010}
 A.S. Khvorostukhin, V.D. Toneev, D.N. Voskresensky, Nucl. Phys. {\bf A845}, 106 (2010).
\bibitem{Plumari2012}
S.Plumari, A. Paglisi, F. Scardina and V. Greco,Phys. Rev.{\bf C86}, 054902 (2012).
 \bibitem{Gorenstein2008}
 M. I. Gorenstein, M. Hauer, O. N. Moroz, Phys.Rev.{\bf C77}, 024911 (2008).
\bibitem{Greiner2012}
 J. Noronha-Hostler, J. Noronha and C. Greiner , Phys. Rev. {\bf C86}, 024913 (2012).
\bibitem{TiwariSrivastava2012}
 S.K. Tiwari, P.K. Srivastava, C.P. Singh, Phys.Rev. {\bf C85}, 014908 (2012).
\bibitem{GhoshMajumder2013}
 S. Ghosh, A. Lahiri, S. Majumder, R. Ray, S. K. Ghosh, Phys. Rev. {\bf C88}, 068201 (2013).
\bibitem{Weise2015}
R. Lang, N. Kaiser, and W. Weise, Eur. Phys. J. {\bf A51}, 127 (2015).
\bibitem{GhoshSarkar2014}
 S. Ghosh, G. Krein, S. Sarkar, Phys.Rev. {\bf C89}, 045201 (2014).
\bibitem{WiranataKoch} 
A. Wiranata, V. Koch and M. Prakash, X.N. Wang, J.Phys.Conf.Ser.{\bf 509}, 012049 (2014). 
\bibitem{WiranataPrakashChakrabarty2012}
 A. Wiranata, M. Prakash and P. Chakraborty, Central Eur.J.Phys. {\bf 10}, 1349-1351 (2012).
\bibitem{Wahba2010}
 A. Tawfik and M. Wahba, Ann. Phys. {\bf 522}, 849-856 (2010).
 
 
\bibitem{Greiner2009}
J. Noronha-Hostler,J. Noronha and C. Greiner, Phys. Rev. Lett.{\bf103}, 172302 (2009).
\bibitem{KadamHM2015}
G. Kadam, H. Mishra, Nuclear Physics {\bf A934}, 133147  (2015).
\bibitem{Kadam2015}
G. Kadam, Mod.Phys.Lett. {\bf A30}, no.10, 1550031  (2015).
\bibitem{Ghoshijmp2014}
S. Ghosh, Int. J. Mod. Phys. {\bf A29}, 1450054 (2014).
\bibitem{Demir2014}
N. Demir and A. Wiranata, J.Phys.Conf.Ser.{\bf 535}, 012018 (2014). 
\bibitem{Ghosh2014}
S. Ghosh, Phys. Rev. {\bf C90}, 025202 (2014).
\bibitem{smash}
J.-B. Rose, J. M. Torres-Rincon, A. Schäfer, D. R. Oliinychenko, and H. Petersen, Phys. Rev. C 97, 055204 (2018).
\bibitem{bamps}
C. Wesp, A. El, F. Reining, Z. Xu, I. Bouras, and C. Greiner, Phys. Rev. C 84, 054911 (2011).
\bibitem{bamps2}
Moritz Greif, Ioannis Bouras, Carsten Greiner, and Zhe Xu, Phys. Rev. D 90, 094014 (2014).
\bibitem{urqmd1}
S. A. Bass et. al, Prog. Part. Nucl. Phys. {\bf 41}, 225 (1998).
\bibitem{GURUHM2015}
G. Kadam, H. Mishra, Phys. Rev. {\bf C92}, 035203 (2015).
\bibitem{ranjitahm}
R. K. Mohapatra, H. Mishra, S. Dash, B. K. Nandi, arXiv:1901.07238.
\bibitem{amanhm1}
P. Singha, A. Abhishek, G. Kadam, S. Ghosh, H. Mishra, J. Phys. {\bf G46}, 015201.
\bibitem{amanhm2}
A. Abhishek, H. Mishra, S. Ghosh, Phys. Rev. {\bf D97}, 014005 (2018).
\bibitem{arpanhm}
J. R. Bhatt, A. Das, H. Mishra, Phys. Rev. {\bf D99}, 014015 (2019).
\bibitem{Rischkemag}
X. Huang, A. Sedrakian, 
D. H. Rischke, Annals Phys. 326 (2011) 3075.
\bibitem{hallhrg}
  A. Das, H. Mishra, R. K. Mohapatra, Phys.Rev. D99 (2019) no.9, 094031.
  \bibitem{hallqgp}
  A. Das, H. Mishra, R. K. Mohapatra, arXiv:1907.05298.
  
  \bibitem{semiconductor}
``Basic Semiconductor Physics'', C. Hamaguchi, Spinger, Spinger-Verlag Berlin Heidelberg 2001, 2010, DOI: 10.1007/978-3-642-03303-2
\bibitem{pairplasma1}
A.Kandus, C. G. Tsagas, Mon. Not. R. Astron. Soc. {\bf 385}, 883-892 (2008).
\bibitem{pairplasma2}
E. G. Blackman, G. B. Field, Phys. Rev. Lett. {\bf 71}, 3481 (1993).
\bibitem{pairplasma3}
N. Bessho, A. Bhattacharjee, Physics of Plasmas {\bf 14}, 056503 (2007).

  \bibitem{landaubook}
 L. P. Pitaevskii, E.M. Lifshitz, Physical Kinetics: Volume 10 (Course of Theoretical Physics).
  \bibitem{sedrakianPRD}
  A. Harutyunyan and A. Sedrakian, Phys. Rev. C94,no. 2, 025805 (2016).
  \bibitem{tuchin3}
  K. Tuchin, J. Phys. G: Nucl. Part. Phys.39(2012)  025010.

  \bibitem{hattori1}
  K. Hattori, D. Satow, Phys.Rev. D94 (2016) no.11, 114032.
  \bibitem{hattori2}
  Koichi Hattori, Shiyong Li, Daisuke Satow, Ho-Ung Yee, Phys.Rev. D95 (2017) no.7, 076008.
  \bibitem{hattori3}
  K. Hattori, Xu-Guang Huang, D. H. Rischke, D. Satow, Phys.Rev. D96 (2017) no.9, 094009.
  \bibitem{vinod1}
  M. Kurian, V. Chandra, Phys.Rev. D99 (2019) no.11, 116018. 
\bibitem{feng2017}
B. Feng, Phys. Rev. {\bf D96}, 036009 (2017).

  
  \bibitem{HRG1}
P. Braun-Munzinger, K. Redlich, J. Stachel, nucl-th/0304013.
\bibitem{HRG2}
 A. Andronic, P. Braun-Munzinger, J. Stachel, Nucl. Phys. {\bf A772}, 167 (2006).
 \bibitem{HRG3}
   P. Braun-Munzinger, D. Magestro, K. Redlich, J. Stachel, Phys. Lett. {\bf B518}, 41 (2001);
  Cleymans, K. Redlich, Phys. Rev.{\bf C60}, 054908 (1999);
   F. Becattini, et al., Phys. Rev.{\bf C64}, 024901 (2001);
 Cleymans, B. Kampfer, M. Kaneta, S. Wheaton, N. Xu, Phys. Rev.{\bf C71}, 054901 (2005);
 A. Andronic, P. Braun-Munzinger, J. Stachel, Phys. Lett. {\bf B673}, 14 (2009).
 
 \bibitem{HRG4}
 R. Dashen, S. Ma, and H. J. Bernstein, Phys. Rev. {\bf187}, 345 (1969).

 \bibitem{HRG5}
R. Dashen and R. Rajaraman, Phys.Rev. {\bf D10}, 694 (1974).

\bibitem{thermodynamicsHRG1}
F. Karsch, K. Redlich, A. Tawfik, Phys.Lett. B571, 67-74 (2003).
\bibitem{thermodynamicsHRG2}
P. Braun-Munzinger, V. Koch, T. Schafer, J. Stachel,  Phys.Rept. 621, 76 (2016).




\bibitem{hrgfluc3}
M. Nahrgang, M. Bluhm, P. Alba, R. Bellwied, C. Ratti, Eur.Phys.J. C75, no.12, 573 (2015).
\bibitem{hrgfluc4}
A. Bhattacharyya, S. Das, S. K. Ghosh, R. Ray, S. Samanta,
Phys.Rev. {\bf C}90, no.3, 034909 (2014).
\bibitem{rmfluc}
R. K. Mohapatra, Phys. Rev. C 99, 024902 (2019).
\bibitem{hrgfluc5}
P. Garg, D.K. Mishra, P.K. Netrakanti, B. Mohanty, A.K. Mohanty,
B.K. Singh, N. Xu,  Phys.Lett. B726, 691-696 (2013).
\bibitem{hrgfluc6}
A. Bazavov et al. Phys.Rev. {\bf D} 86, 034509 (2012).
\bibitem{hrgfluc7}
V.V. Begun, M. I. Gorenstein, M. Hauer, V.P. Konchakovski, O.S. Zozulya,
Phys.Rev. {\bf C} 74, 044903 (2006).
\bibitem{stockerRischke}
D.H.Rischke, M.I.Gorenstein, H.Stocker, W.Greiner; Z. Phys. C 51,485-489 (1991)
\bibitem{albrightkapusta}
M. Albright, J.I. Kapusta, Phys.Rev. C93 (2016) no.1, 014903.
\bibitem{paramitahm2016}
P. Deb, G. Kadam, H. Mishra, Phys. Rev. {\bf D 94},094002 (2016).

 \bibitem{thermoelec1}
  A. Cantarero and F. X. Alvarez, “Thermoelectric Effects: Semiclassical and Quantum Approaches from the BoltzmannTransport Equation”, published in Lect. Notes in NanoscaleScience and Technology, vol 16 “Nanoscale Thermoelectrics”,edited by X. Wang and Z. M. Wang.
  \bibitem{thermoelec2}G.S.Nolas, J. Sharp and H. J. Goldsmid, “Thermoelectrics: Basic Principles and New Materials Developments”, Springerseries in Materials Science, vol 45.
\bibitem{lataguruhm}
G. Kadam, H. Mishra, L. Thakur, Phys. Rev. {\bf D 98}, 114001 (2018). 
\bibitem{sabya1}
  J. Dey, S. Satapathy, P. Murmu, S. Ghosh, arXiv:1907.11164.
  \bibitem{sabya2}
  J. Dey, S. Satapathy, A. Mishra, S. Paul, S. Ghosh, arXiv:1908.04335.
  \bibitem{sabya3}
  S. Satapathy, S. Paul, A. Anand, R. Kumar, S. Ghosh, arXiv:1908.04330.
  \bibitem{payal}
  P. Mohanty, A. Dash, V. Roy, Eur.Phys.J. A55, 35 (2019).  
\bibitem{HRGMuller}
A. Majumder and B. Muller, Phys. Rev. Lett {\bf 105},252002 (2010).
\bibitem{gondologelmini}
P. Gondolo and G. Gelmini, Nucl. Phys. {\bf B360}, 145 (1991).
\bibitem{pdg}
C. Amsler et al.[Particle Data Group], Phys. Lett. {\bf B 667}, 1 (2008).
\bibitem{kapustaAlbr}
M. Albright, J. Kapusta, C. Young, Phys.Rev. C {\bf 90} no.2, 024915 (2014). 
\bibitem{hmgururadius1}
P. Braun-Munzinger, I. Heppe, J. Stachel, Phys. Lett. {\bf B465}, 15 (1999).
  
 
  
\end{thebibliography}
\end{document}